\documentclass[sigconf]{acmart}

\usepackage{booktabs} 
\usepackage{paralist}

\newif\ifarxiv
\arxivtrue

\copyrightyear{2017} 
\acmYear{2017} 
\setcopyright{acmlicensed}
\acmConference{SoCC '17}{September 24--27, 2017}{Santa Clara, CA,
USA}\acmPrice{15.00}\acmDOI{10.1145/3127479.3132244}
\acmISBN{978-1-4503-5028-0/17/09}

\usepackage[utf8]{inputenc}
\usepackage[noend,ruled,linesnumbered]{algorithm2e}
\usepackage{tabu}
\usepackage{url}
\urldef\acmdoiurl\url{https://doi.org/10.1145/3127479.3132244}

\newcommand\blfootnote[1]{%
  \begingroup
  \renewcommand\thefootnote{}\footnote{#1}%
  \addtocounter{footnote}{-1}%
  \endgroup
}

\usepackage[caption=false,font=scriptsize]{subfig}

\newcommand\fixme[1]{}

\begin{document}

\title[SLO-aware Colocation of Data Center Tasks Based on Inst. Proc. Requirements]{SLO-aware Colocation of Data Center Tasks Based on Instantaneous Processor Requirements}

\author{Pawel Janus}
\affiliation{%
  \institution{Institute of Informatics, University of Warsaw}
  \streetaddress{Banacha 2}
  \city{Warsaw}
  \country{Poland}
}
\email{pj320664@students.mimuw.edu.pl}

\author{Krzysztof Rzadca}
\affiliation{%
  \institution{Institute of Informatics, University of Warsaw}
  \streetaddress{Banacha 2}
  \city{Warsaw}
  \country{Poland}
}
\email{krz@mimuw.edu.pl}

\begin{abstract}
  In a cloud data center, a single physical machine simultaneously executes dozens of highly heterogeneous tasks.
  Such colocation results in more efficient utilization of machines, but, when tasks' requirements exceed available resources, some of the tasks might be throttled down or preempted.
  We analyze version 2.1 of the Google cluster trace that shows short-term (1 second) task CPU usage.
  Contrary to the assumptions taken by many theoretical studies, we demonstrate that the empirical distributions do not follow any single distribution.
  However, high percentiles of the total processor usage (summed over at least 10 tasks) can be reasonably estimated by the Gaussian distribution.
  We use this result for a probabilistic fit test, called the Gaussian Percentile Approximation (GPA), for standard bin-packing algorithms.
  To check whether a new task will fit into a machine, GPA checks whether the resulting distribution's percentile corresponding to the requested service level objective, SLO is still below the machine's capacity.
  In our simulation experiments, GPA resulted in colocations exceeding the machines' capacity with a frequency similar to the requested SLO.
\end{abstract}



\maketitle

\section{Introduction}
Quantity or quality? When a cloud provider colocates more tasks on a machine, the infrastructure is used more efficiently.
\ifarxiv
\blfootnote{Author's version of a paper published in ACM SoCC ’17.
    The Definitive Version of Record is available at ACM Digital Library at
  \acmdoiurl
}
\fi
In the short term, the throughput increases. In the longer term, packing more densely reduces future investments in new data centers. However, if the tasks' requirements exceed available resources, some of the tasks might be throttled down or preempted, affecting their execution time or performance. 

A standard solution to the quantity-quality dilemma is the Service Level Agreement (SLA): the provider guarantees a certain quality level, quantified by the Service Level Objective (SLO, e.g., a VM with an efficiency of one x86 core of 2Ghz) and then maximizes the quantity. However, rarely is the customer workload using the whole capacity the whole time. The provider has thus a strong incentive to oversubscribe (e.g., to rent 11 single-core virtual machines all colocated on a single 10-core physical CPU), and thus to change the quality contract to a probabilistic one (a VM with a certain efficiency at least 99\% of the time).

In this paper, we show how to maintain a probabilistic SLO based on \emph{instantaneous} CPU requirements of tasks from a Google cluster~\cite{clusterdata:Reiss2011}. Previous approaches packed tasks based on the \emph{maximum} or the \emph{average} CPU requirements.
The maximum corresponds to no oversubscription, while the average can severely overestimate the quality of service (QoS). Consider the following example with two machines with CPU capacity of $1.0$ each and three tasks $t_1, t_2, t_3$. Tasks $t_1$ and $t_2$ are stable with constant CPU usage of $0.5$. In contrast, task $t_3$'s CPU usage varies: assume that it is drawn from a uniform distribution over $[0, 1]$.
If the resource manager allocates tasks based only on their average CPU usage, $t_3$, having mean $0.5$, can end up packed to a machine shared with either $t_1$ or $t_2$; thus, this machine will be overloaded half of the time. An alternative allocation, in which $t_1$ and $t_2$ share a machine, and $t_3$ has a dedicated machine, uses the same number of machines, and has no capacity violations.

To the best of our knowledge, until recently, all publicly available data on tasks' CPU usage in large systems had a very low time resolution. The Standard Workload Format~\cite{chapin_benchmarks_1999} averages CPU usage over the job's entire runtime. The Google cluster trace~\cite{clusterdata:Wilkes2011, clusterdata:Reiss2011} in versions 1.0 and 2.0 reports CPU usage for tasks (Linux containers) averaged over 5-minute intervals (the \texttt{mean CPU usage rate} field). Relying on this field, as our toy example shows, can result in underestimation of the likelihood of overload.

The Google Cluster Trace in version 2.1 extended the \texttt{resource usage} table with a new column, the \texttt{sampled CPU usage}.
The \texttt{resource usage} table contains a single record for each 5 minutes runtime of each task.
The \texttt{sampled CPU usage} field specifies the CPU usage of a task averaged over \emph{a single second} randomly chosen from these 5 minutes.
Different tasks on the same machine are not guaranteed to be sampled at the same moment; and, for a task, the sampling moment is not the same in different 5-minute reporting periods.
In contrast, another field, the \texttt{mean CPU usage rate}, shows the CPU usage averaged over the whole 5 minutes.
Later on, to avoid confusion, we refer to \texttt{sampled CPU usage} as the \emph{instantaneous (inst)} CPU usage, and to \texttt{mean CPU usage rate} as the \emph{(5-minute) average (avg)} CPU usage.


As we show in this paper, the data on instantaneous CPU usage brings a new perspective on colocation of tasks. First, it shows how variable cloud computing tasks are in shorter time spans. Second, it is one of the few publicly available, realistic data sets for evaluating stochastic bin packing algorithms. 
The contributions of this paper are as follows:
\begin{itemize}
\item The instantaneous usage has a significantly higher variability than the previously used 5-minute averages (Section~\ref{sec:data-analysis}).
  For longer running tasks, we are able to reconstruct the complete distribution of the requested CPU. We show that tasks' usage do not fit any single  distribution. 
  However, we demonstrate that we are able to estimate high percentiles of the total CPU demand when 10 or more tasks are colocated on a single machine.
\item We use this observation for a test, called the Gaussian Percentile Approximation (GPA), that checks whether a task will fit into a machine on which other tasks are already allocated (Section~\ref{sec:gpa-algorithm}). Our test uses the central limit theorem to estimate parameters of a Gaussian distribution from means and standard deviations of the instantaneous CPU usage of colocated tasks. Then it compares the machine capacity with a percentile of this distribution corresponding to the requested SLO. According to our simulations (Section~\ref{sec:experiments}), colocations produced by GPA have QoS similar to the requested SLO.
\end{itemize}

The paper is organized as follows. To guide the discussion, we present the assumptions commonly taken by the stochastic bin packing approaches and their relation to data center resource management in Section~\ref{sec:definition}. Section~\ref{sec:data-analysis} analyzes the new data on the instantaneous CPU usage. Section~\ref{sec:gpa-algorithm} proposes GPA, a simple bin packing algorithm stemming from this analysis. Section~\ref{sec:experiments} validates GPA by simulation. Section~\ref{sec:rel-work} presents related work.

\section{Problem Definition}\label{sec:definition}
Rather than trying to mimic the complex mix of policies, algorithms and heuristics used by real-world data center resource managers~\cite{verma2015large,schwarzkopf_omega:_2013}, we focus on a minimal algorithmic problem that, in our opinion, models the core goal of a data~center resource manager: collocation, or VM consolidation i.e. which tasks should be colocated on a single physical machine and how many physical machines to use.
Our model focuses on the crucial quantity/quality dilemma faced by the operator of a datacenter: increased oversubscription results in more efficient utilization of machines, but decreases the quality of service, as it is more probable that machine's resources will turn out to be insufficient.
We model this problem as stochastic bin packing i.e., bin packing with stochastically-sized items~\cite{kleinberg_allocating_2000,goel1999stochastic}. We first present the problem as it is defined in~\cite{kleinberg_allocating_2000,goel1999stochastic}, then we discuss how appropriate are the typical assumptions taken by theoretical approaches for the data center resource management.  

In stochastic bin packing, we are given a set $S$ of $n$ items (tasks) $\{X_1, \dots, X_n\}$. $X_i$ is a random variable describing task $i$'s resource requirement. We also are given a threshold $c$ on the amount of resources available in each node (the capacity of the bin); and a maximum admissible overflow probability $\varrho$, corresponding to a Service Level Objective (SLO). The goal is to partition $S$ into a minimal number $m$ of subsets $S_1, \dots, S_m$, where a subset $S_j$ corresponds to tasks placed on a (physical) machine $j$. The constraint is that, for each machine, the probability of exceeding its capacity is smaller than the SLO  $\varrho$, i.e., $Pr[ \sum_{i: X_i \in S_j} X_i > c ] < \varrho$.

Stochastic bin packing assumes that there is \emph{no notion of time}: all tasks are known and ready to be started, thus all tasks should be placed in bins. While resource management in a datacenter typically combines bin packing and scheduling~\cite{verma2015large,schwarzkopf_omega:_2013, DBLP:conf/ipps/TangLRC16}, we assume that the schedule is driven by higher-level policy decisions and thus beyond the optimization model.
Moreover, even if the schedule can be optimized, eventually the tasks have to be placed on machines using a bin-packing-like approach, so a better bin-packing method would lead to a better overall algorithm.

Stochastic bin packing assumes that the items to pack are \emph{one-dimensional}. Resource usage of tasks in a data~center can be characterized by at least four measures~\cite{clusterdata:Reiss2012b,stillwell2012virtual}: CPU, memory, disk and network bandwidth. One-dimensional packing algorithms can be extended to multiple dimensions by vector packing methods~\cite{lee_validating_2011,stillwell2012virtual}.

Stochastic bin packing assumes that tasks' resource requirements are stochastic (random) \emph{variables}, thus they are \emph{time-invariant} (in constrast to stochastic \emph{processes}).
The analysis of the previous version of the trace~\cite{clusterdata:Reiss2012b} concludes that for most of the tasks the hour to hour ratio of the average CPU usage does not change significantly.  This observation corresponds to an intuition that datacenter tasks execute similar workload over longer time periods.
Moreover, as the instantaneous usage is just a single 1-second sample from a 5-minute interval, any short term variability cannot be reconstructed from the data. For instance, consider a task with an oscillating CPU usage rising as a linear function from 0 to 1 and then falling with the same slope back to 0. If the period is smaller than the reporting period (5 minutes), the ``sampled CPU usage'' would show values between 0 and 1, but without any order; thus, such a task would be indistinguishable from a task that draws its CPU requirement from a uniform distribution over $[0,1]$.
We validate the time-invariance assumption in Section~\ref{subsec:stationary-proc}.

To pack tasks, we need information about their sizes. Theoretical approaches commonly  assume \emph{clairvoyance}, i.e., perfect information~\cite{goel1999stochastic,stillwell2012virtual,wang_consolidating_2011,chen2011effective}. In clairvoyant stochastic bin packing, while the exact sizes---realizations---are unknown, the distributions $X_i$ are known.
We test how sensitive the proposed method is to available information in Section~\ref{sec:exp-clairvoyance}, where we provide only a limited fraction of measurements to the algorithms. Clearly, a data~center resource manager is usually unable to test a task's usage by running it for some time before allocating it.
However, a task's usage can be predicted by comparing the task to previously submitted tasks belonging to the same or similar jobs (similarity can be inferred from, e.g., user's and job's name).
Our limited clairvoyance simulates varying quality of such predictions.
Such prediction is orthogonal to the main results of this paper. 
We do not rely on user supplied information, such as the declared maximum resource usage, as these are rarely achieved~\cite{clusterdata:Reiss2012b}.
In contrast to standard stochastic bin packing, our solution does not use the distributions $X_i$ of items' sizes; it only requires two statistics, the mean $\mu_i(X_i)$ and the standard deviation $\sigma_i(X_i)$.

Algorithms for stochastic bin packing typically assume that the items' distributions $\{ X_i \}$ are \emph{independent}. In a data~center, a job can be composed of many individual tasks; if these tasks are, e.g., instances of a large-scale web application, their resource requirements can be correlated (because they all follow the same external signal such as the popularity of the website). 
If correlated tasks are placed on a single machine, estimations of, for instance, the mean usage as the sum of the task's means are inexact.
However, in a large system serving many jobs,
the probability that a machine executes many tasks from a single job is relatively small (with the exception for data dependency issues~\cite{chowdhury_efficient_2015}).
Thus, a simple way to extend an algorithm to handle correlated tasks is not to place them on the same machine (CBP,~\cite{verma_server_2009}).
While we acknowledge that taking into account correlations is an important direction of future work, the first step is to characterize how frequent they are; and analyses of the version 2.0 of the trace~\cite{clusterdata:Reiss2012b,clusterdata:Di2013} did not consider this topic.




While a typical data~center executes tasks of different importance (from critical production jobs to best-effort experiments), stochastic bin packing assumes that all tasks have the same priority/importance. Different priorities can be modeled as different requested SLOs; simultaneously guaranteeing various SLOs for various groups of colocated tasks is an interesting direction of future work.

We also assume that all machines have equal capacities (although we test the impact of different capacities in Section~\ref{sec:exp-varied-capacities}). 

Finally, we assume that exceeding the machine's capacity is undesirable, but not catastrophic. Resources we consider are rate-limited, such as CPU, or disk/network bandwidth, rather than value-limited (such as RAM). If there is insufficient CPU, some tasks slow down; on the other hand, insufficient RAM may lead to immediate preemption or even termination of some tasks.

\section{Characterization of instantaneous CPU usage}\label{sec:data-analysis}
In this section we analyze \texttt{sampled CPU usage}, which we call the instantaneous (inst) CPU usage, introduced in version 2.1 of the Google trace~\cite{clusterdata:Reiss2011}. We refer to~\cite{clusterdata:Reiss2012b,clusterdata:Di2013} for the analysis of the previous version of the dataset.

We use the following notation. We denote by $T_i$ the number of records about task $i$ in the \texttt{resource usage} table (thus, effectively, task's $i$ duration as counted by 5-minute intervals). We denote the $t$-th value of task $i$ \emph{instantaneous (inst) usage} as $x_i(t)$; and the $t$-th value of task $i$ \emph{5-minute average usage} as $y_i(t)$.
We reserve the term \emph{mean} for a value of a statistic $\bar{x_i}$ computed from a (sub)sample, e.g., for the whole duration $(x_i(1), \dots, x_i(T_i))$, $\bar{x_i}=\frac{1}{T_i} \sum_{t=1}^{T_i} x_i(t)$.
We denote by $X_i$ the empirical distribution generated from $(x_i(1), \dots, x_i(T_i))$.


\subsection{Data preprocessing}\label{sec:data-preprocessing}
We first discard all failing tasks as our goal is to characterize a task's resource requirements during its complete execution (in our future work we plan to take into account also the resource requirements of these failing tasks).
We define task as failing if it contains at least one of \texttt{EVICT(2)}, \texttt{FAIL(3)}, \texttt{KILL(5)}, \texttt{LOST(6)} events in the events table.
We then discard $209\,940$ tasks (1.2\% of all tasks in the trace)
that show zero instantaneous usage for their entire duration: these tasks correspond to measurement errors, or truly non-CPU dependent tasks, which have thus no impact on the CPU packing we want to study.

We replace 13 records of average CPU usage higher than 1 by the corresponding instantaneous usage (no instantaneous usage records were higher than 1). The trace normalizes both values to 1 (the highest total node CPU capacity in the cluster). Thus, values higher than 1 correspond to measurement errors (note that these 13 records represent a marginal portion of the dataset).


Task lengths differ: $16\,055\,428$ (95\% of all) tasks are shorter than 2 hours and thus have less than 24 CPU measurements.
We partition the tasks into two subsets, the \emph{long} tasks (2 hours or longer) and the remaining \emph{short} tasks. We analyze only the \emph{long} tasks.
We do not consider the short tasks,
as, first, they account for less than 10\% of the overall utilization of the cluster~\cite{clusterdata:Reiss2012b},
and, second, the shorter the task, the less measurements we have and thus the less reliable is the empirical distribution of the instantaneous usage (see Section~\ref{sec:data-analysis-sampling}).


Finally, some of our results (normality tests, percentile predictions, experiments in Section~\ref{sec:experiments}) rely on repeated sampling of instantaneous and average CPU usage of tasks. For such experiments, we generate a random sample of $N=100\,000$ long tasks. 
For each task from the sample, we generate and store $R=10\,000$ realizations of both instantaneous and average CPU usage.
The instantaneous realizations are generated as follows (averages are generated analogously).
From $(x_i(1), \dots, x_i(T_i))$, we create an empirical distribution (following our assumption of time invariance). We then generate $R$ realizations of a random variable from this distribution. Such representation allows us to have CPU usage samples of equal length independent of the actual duration $T_i$ of the task. Moreover, computing statistics of the total CPU usage with such long samples is straightforward: e.g., to get samples of the total CPU usage for 3 tasks colocated on a single node, it is sufficient to add these tasks' sampled instantaneous CPU usage, i.e., to add 3 vectors, each of $10\,000$ elements.
Our data is available at \url{http://mimuw.edu.pl/~krzadca/sla-colocation/}.


\subsection{Validation of Instantaneous Sampling}\label{sec:data-analysis-sampling}


\begin{figure}[!tb]
  \includegraphics[width=\columnwidth]{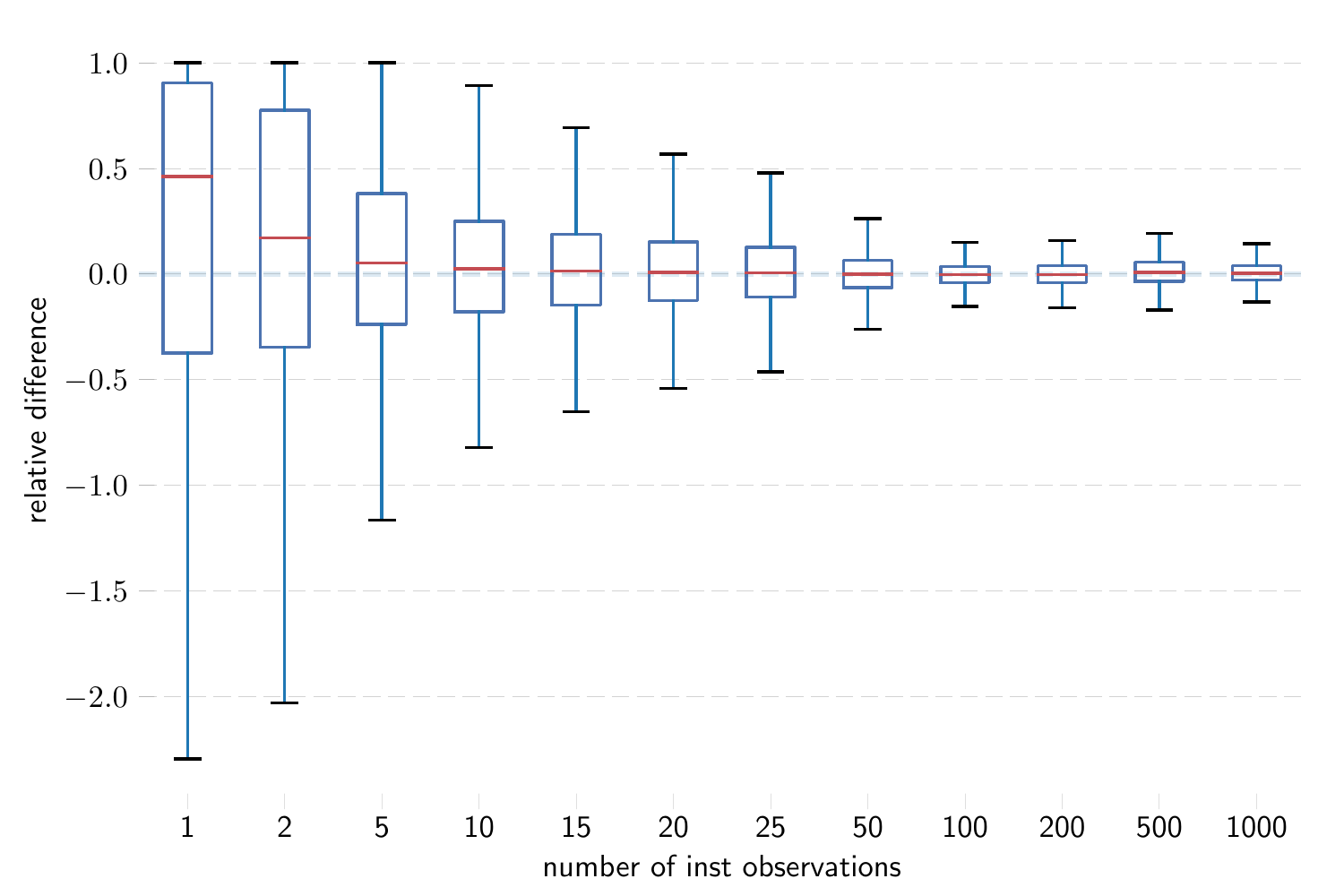}
  \caption{Distributions of the relative differences
    ($(\bar{y_i}-\bar{x_i}^{(k)})/\bar{y_i}$) 
    between the means computed from 5-minute average $y_i$ and instantaneous $x_i$
    CPU usage as a function of the number of instantaneous samples $k$ for all tasks at least 2 hours long.
Here, and in the remaining boxplots, the line inside the box denotes the median; the box spans between the first and the third quartile (the interquartile range, IQR); and the whiskers extend to the most extreme data point within 1.5 $\times$ IQR.}
      \label{fig:means_comparison}
\end{figure}

We start by evaluating
whether tasks' instantaneous samples are representative of their true usage, i.e., whether the method used to produce  instantaneous  data was unbiased. While we don't know the true usage, we have an independent measure, the 5-minute averages. Our hypothesis is that the mean of the instantaneous samples should converge to the mean of the 5-minute average samples.
Figure~\ref{fig:means_comparison} shows the distribution of the relative difference of means as a function of the number of samples.
For a task $i$ we compute the mean of the average CPU usage $\bar{y_i}=\frac{1}{T_i} \sum_t y_i(t)$ (taking into account all measurements $y_i(t)$ during the whole duration of the task). We then compute the mean of a given number $k$ of instantaneous CPU usage $\bar{x_i}^{(k)}=\frac{1}{k} \sum_{t \in S_k} x_i(t)$ ($k \in \{ 1, 2, 5, \dots, 500, 1\,000 \}$, $S_k$ is a randomly chosen subset of $\{1, \dots, T_i\}$ of size k).
For each $k$ independently, we compute the statistics over all tasks having at least $k$ records: thus $k=1$ shows a statistics over all \emph{long} tasks, and $k=1\,000$ over tasks longer than 83 hours.

The figure shows that, in general, the method used to obtain the instantaneous data is unbiased. From approx. 15 samples onwards, the interquartile range, IQR, is symmetric around 0. The more samples, the smaller is the variability of the relative difference, thus, the closer are means computed from the instantaneous and the average data.


\begin{figure}[!tb]
  \includegraphics[width=\columnwidth]{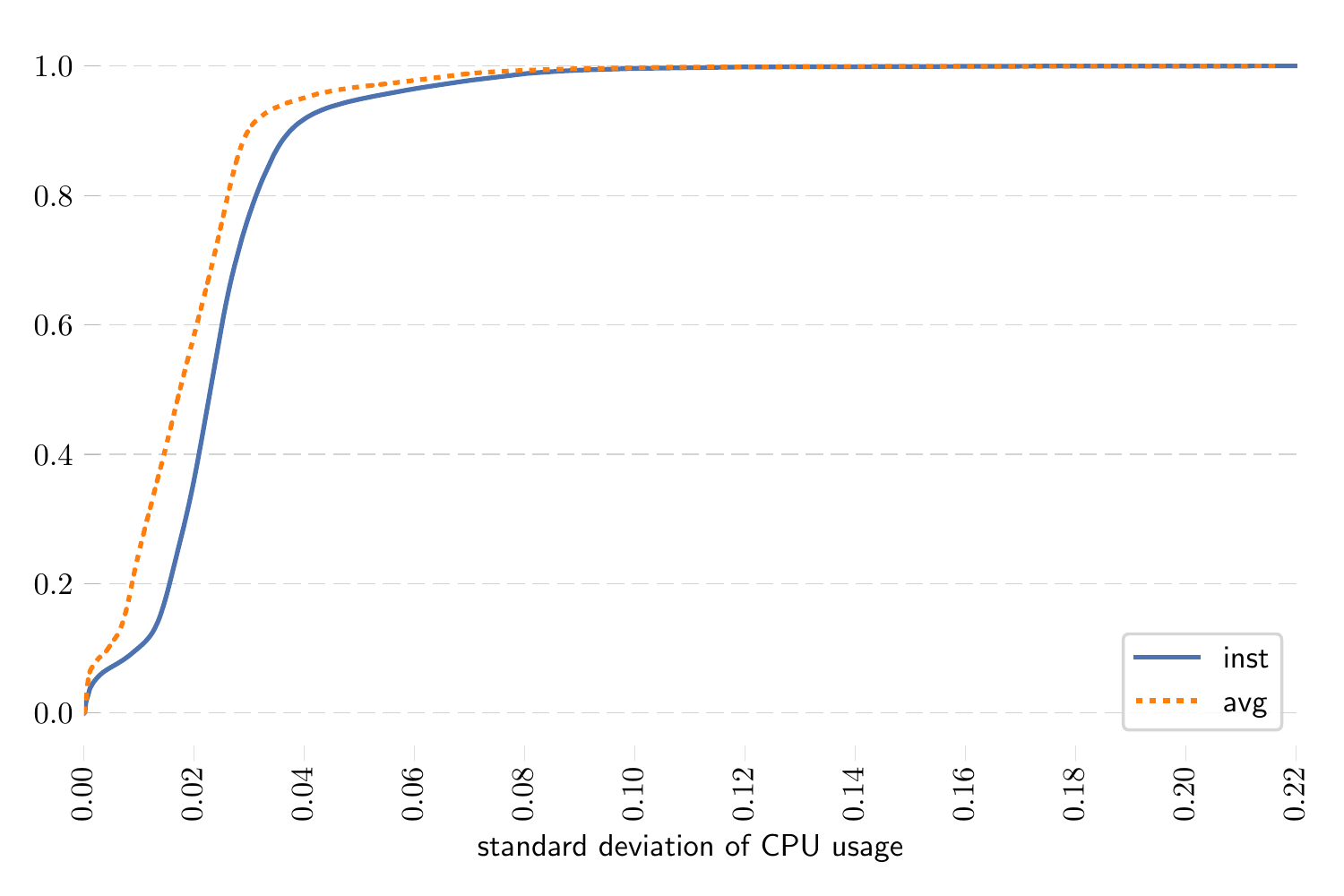}
  \caption{CDF of the distribution of standard deviations of CPU usage for all tasks at least 2 hours long.}
  \label{fig:std-cdf}
\end{figure}

\begin{figure*}
\centering
  \subfloat[inst]{{\includegraphics[width=.33\textwidth]{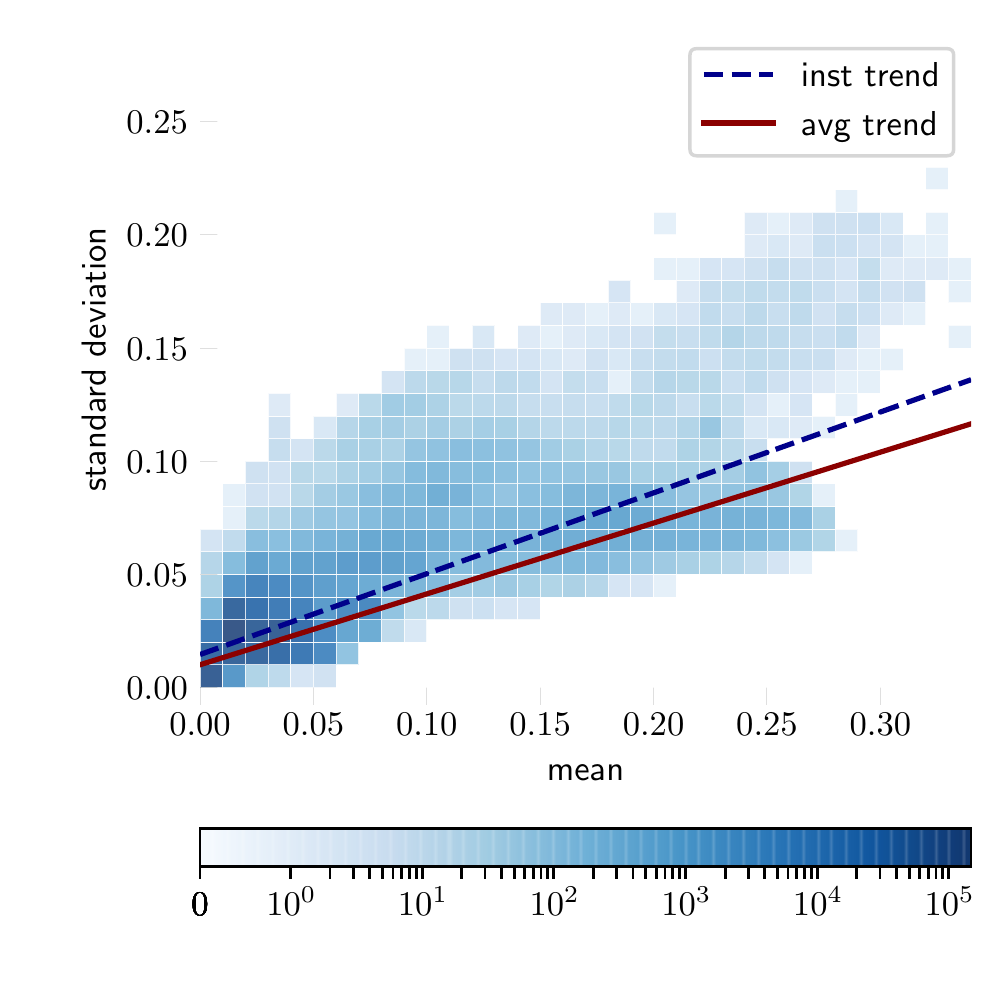}}}%
  \subfloat[avg]{{\includegraphics[width=.33\textwidth]{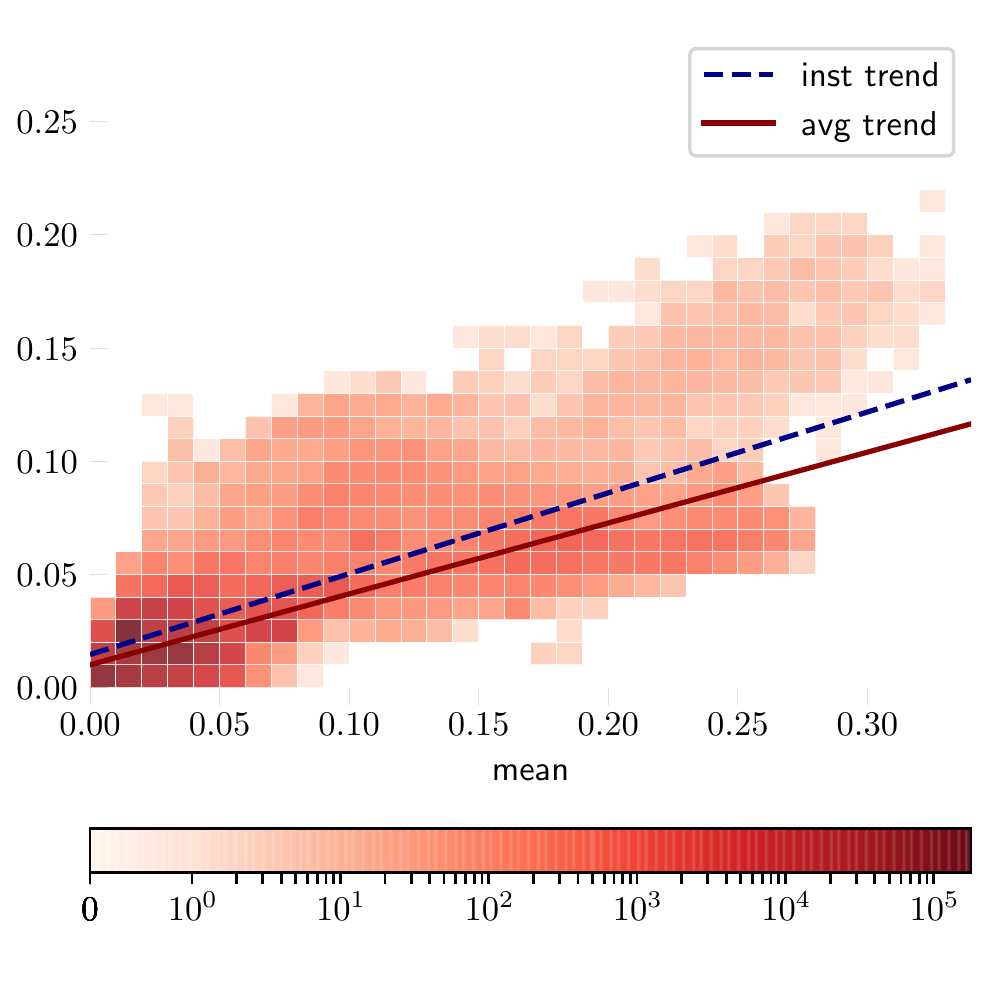}}}%
  \subfloat[inst-avg]{{\includegraphics[width=.33\textwidth]{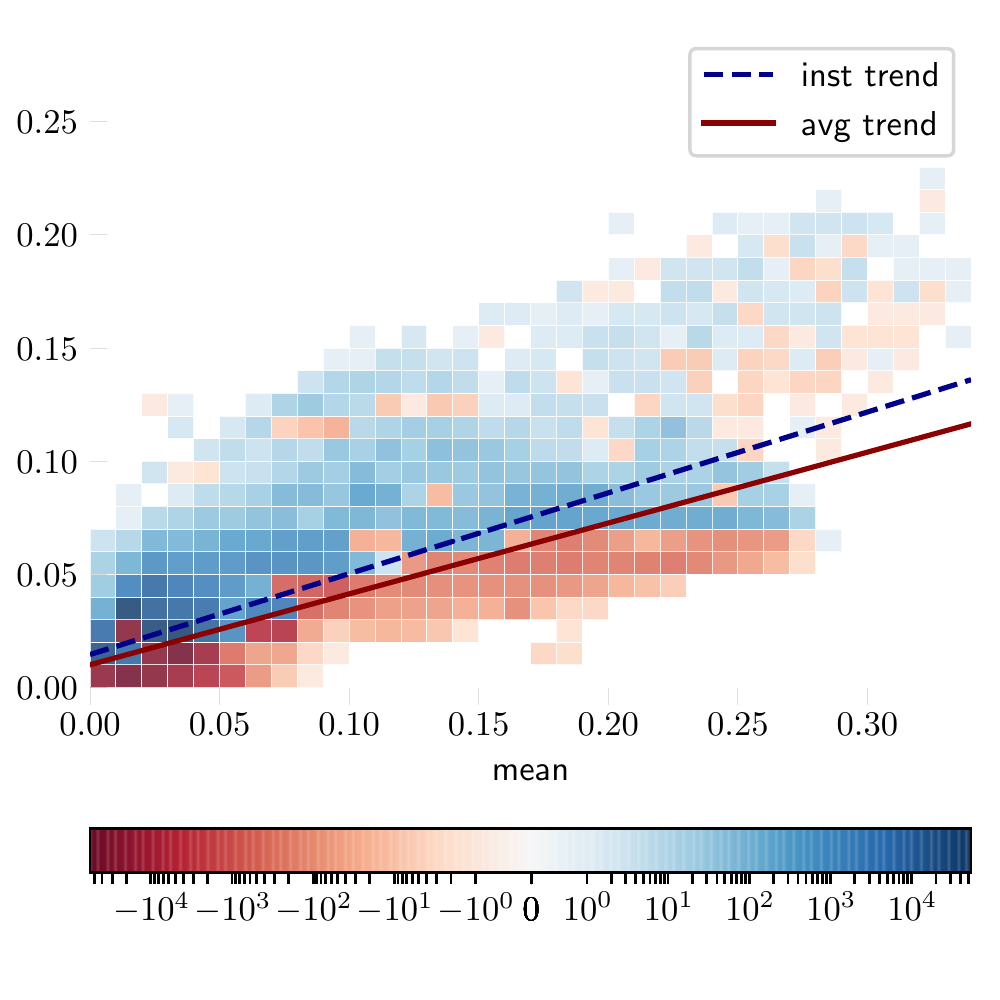}}}%
  \caption{Heatmaps showing standard deviations of CPU usage as a function of means for long tasks. Figure~(c) highlights the differences between (a) and (b): blue areas correspond to (mean $\times$ std) parameters matching more inst than avg samples.}
  \label{fig:std-vs-mean}
\end{figure*}

\subsection{Variability of instantaneous and of average usage}\label{sec:data-analysis-variability}

Next, we characterize the variability of the instantaneous usage as characterized by standard deviations of instantaneous $\sigma_{inst}(i)$ and average $\sigma_{avg}(i)$ usage.
Figure~\ref{fig:std-cdf} shows the CDFs of the standard deviations across all long tasks from the trace. Instantaneous usage is more variable than 5-minute averages. Furthermore, as Figure~\ref{fig:std-vs-mean} shows, standard deviations depend on the mean CPU usage: the higher task's mean CPU usage, the higher its standard deviation: compared to the avg trend line (linear regression) the inst trend line has both steeper slope (0.36 vs. 0.31) and higher intercept (0.015 vs. 0.010).


\subsection{Time invariance}\label{subsec:stationary-proc}
We now test our assumption that the instantaneous loads are drawn from a random distribution that does not depend on time.
For each of the long tasks, we divide observations into windows of consecutive $\Delta=12$ records (a window corresponds to 1 hour): a single window is thus $( x_i(k\Delta+1), x_i(k\Delta+2), \dots, x_i(k\Delta + \Delta-1) )$ (where $k$ is a non-negative integer).
We then compare the distributions of two windows picked randomly (for two different $k$ values, $k_1$ and $k_2$; $k_1$ and $k_2$ differ between tasks). 
Our null hypothesis is that these samples are generated by the same distribution.
\fixme{krz to pawel: read and edit the following}
In contrast, if there is a stochastic \emph{process} generating the data (corresponding to, e.g., daily or weekly usage patterns), with high probability the two distributions would differ (for a daily usage pattern, assuming a long-running task, the probability of picking two hours 24-hours apart is 1/24).

To validate the hypothesis, we perform a Kolmogorov-Smirnov test.
For roughly $30\%$ of tasks the test rejects our hypothesis at the significance level of $5\%$ (the results for $\Delta=24$ and $\Delta=36$ are similar). Thus for roughly $30\%$ of tasks the characteristics of the instantaneous CPU usage changes in time.
On the other hand, the analysis of the average CPU usage~\cite{clusterdata:Reiss2012b} shows that the hour-to-hour variability of individual tasks is small (for roughly 60\% of tasks weighted by their duration, the CPU utilization changes by less than 15\%).
We will further investigate these changing tasks in future work.


\subsection{Variability of individual tasks}

\begin{figure*}[!tb]
  \centering
  \subfloat[$(1.42\%)$]{{\includegraphics[width=.5\columnwidth]{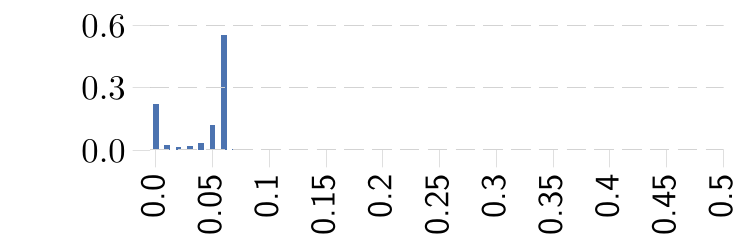}}}%
  \subfloat[$(5.47\%)$]{{\includegraphics[width=.5\columnwidth]{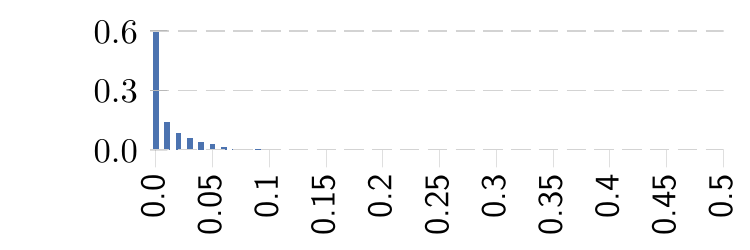}}}%
  \subfloat[$(17.62\%)$]{{\includegraphics[width=.5\columnwidth]{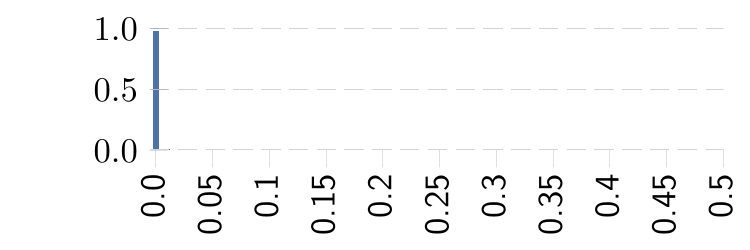}}}%
  \subfloat[$(3.11\%)$]{{\includegraphics[width=.5\columnwidth]{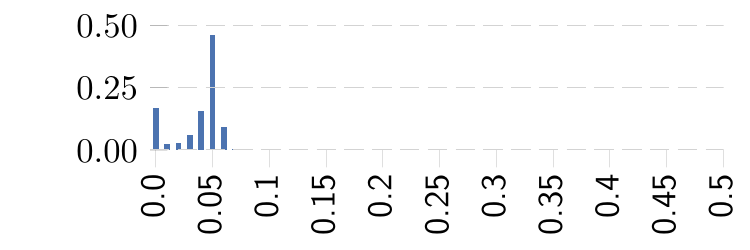}}}%
  \\
  \subfloat[$(4.27\%)$]{{\includegraphics[width=.5\columnwidth]{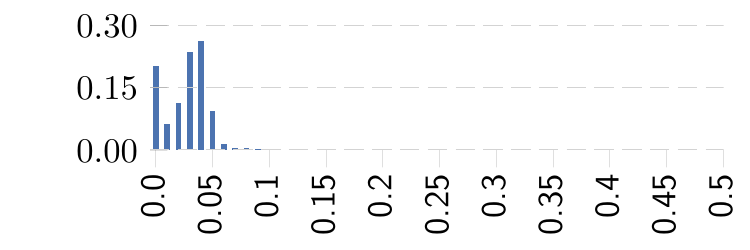}}}%
  \subfloat[$(13.70\%)$]{{\includegraphics[width=.5\columnwidth]{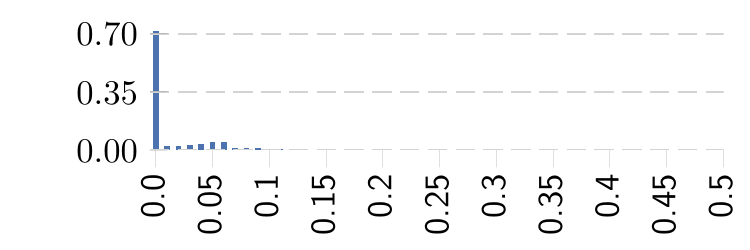}}}%
  \subfloat[$(3.77\%)$]{{\includegraphics[width=.5\columnwidth]{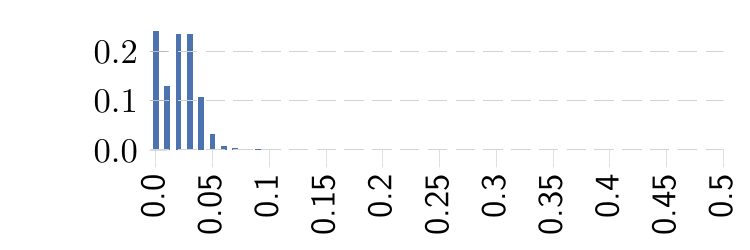}}}%
  \subfloat[$(2.27\%)$]{{\includegraphics[width=.5\columnwidth]{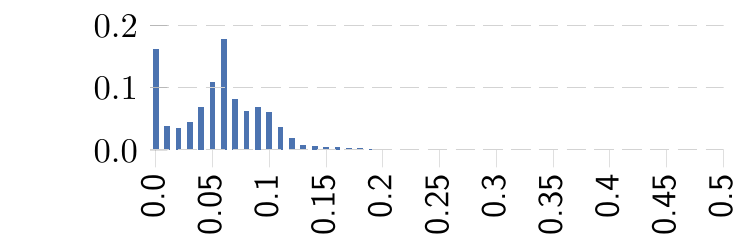}}}%
  \\
  \subfloat[$(2.93\%)$]{{\includegraphics[width=.5\columnwidth]{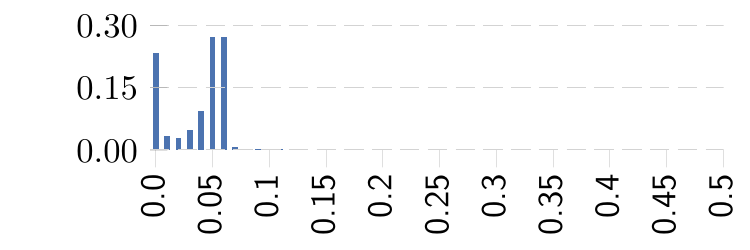}}}%
  \subfloat[$(3.48\%)$]{{\includegraphics[width=.5\columnwidth]{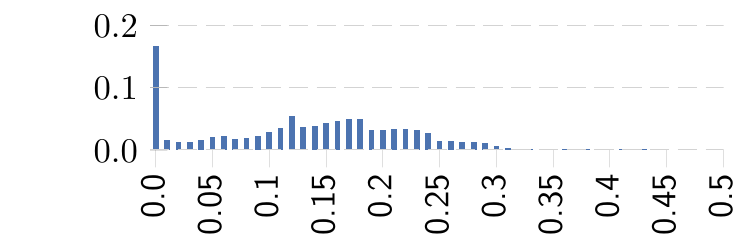}}}%
  \subfloat[$(15.22\%)$]{{\includegraphics[width=.5\columnwidth]{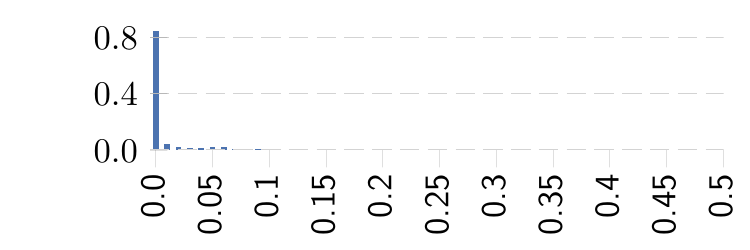}}}%
  \subfloat[$(4.38\%)$]{{\includegraphics[width=.5\columnwidth]{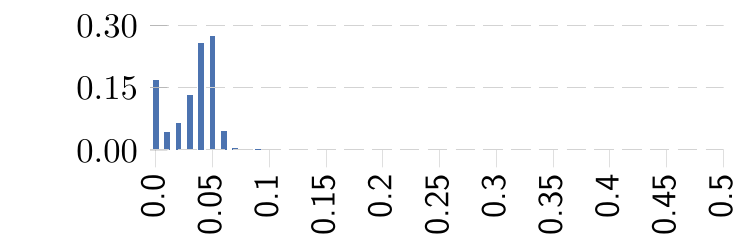}}}%
  \\
  \subfloat[$(6.96\%)$]{{\includegraphics[width=.5\columnwidth]{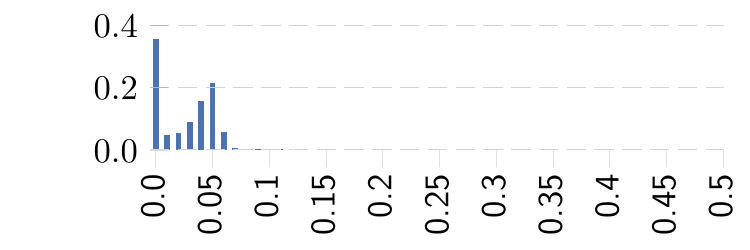}}}%
  \subfloat[$(5.77\%)$]{{\includegraphics[width=.5\columnwidth]{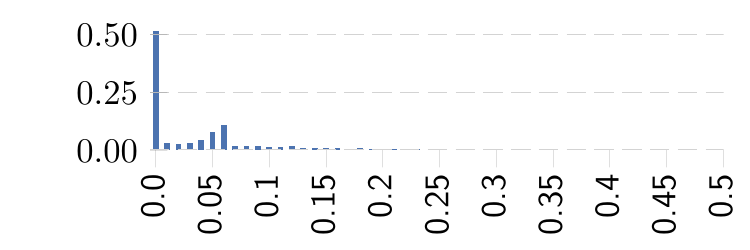}}}%
  \subfloat[$(6.54\%)$]{{\includegraphics[width=.5\columnwidth]{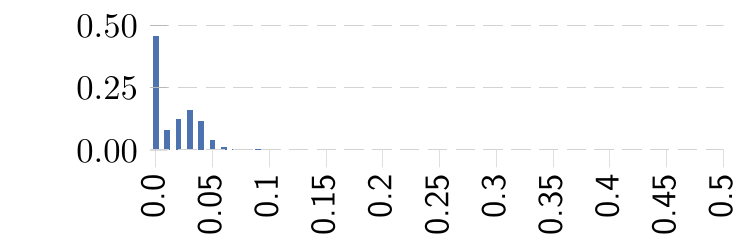}}}%
  \subfloat[$(3.09\%)$]{{\includegraphics[width=.5\columnwidth]{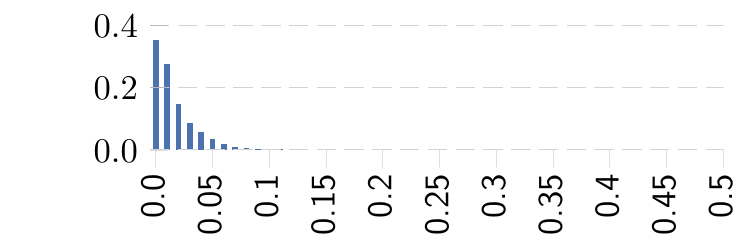}}}%
  \caption{Typical distributions of task's instantaneous CPU usage. Each sub-figure corresponds to a center of one of the 16 clusters produced by the k-means clustering algorithm. 
X---instantaneous usage (cut to $[0,0.50]$); Y---\% share of occurrences (ranges differ between plots).}
  \label{fig:clusters-inst}
\end{figure*}

Many theoretical approaches (e.g.,~\cite{wang_consolidating_2011, goel1999stochastic, zhang_sla_2014}) assume that items' sizes all follow a specific, single distribution (Gaussian, exponential, etc.). In contrast, we discovered that the distributions of instantaneous loads in the Google trace vary significantly among tasks.

To characterize the common types of distributions of roughly $800\,000$ long tasks, we clustered tasks' empirical distributions. Our method is the following.
First, we generate histograms representing the distributions.
We set the granularity of the histogram to  $0.01$.
Let $h_i$ be the histogram of task $i$, a 100-dimensional vector.
$h_i[k]$ (with $0 \leq k \leq 99$)
is the likelihood that an instantaneous usage sample falls between $k/100$ and $(k+1)/100$,
$h_i[k] = | \{ x_i(t) : k/100 \leq x_i(t) < (k+1)/100 \} | / T_i$.

Then, we use the k-means algorithm~\cite{hartigan1975clustering} with the Euclidean distance metric on the set of histograms $\{ h_i \}$. 
The clustering algorithm treats each task as a 100-dimensional vector.
To compute how different tasks $i$ and $j$ are, the algorithm computes the Euclidean distance between $h_i$ and $h_j$.
Typically k-means is not considered an algorithm robust enough for handling high-dimensional data. However, a great majority of our histograms are 0 beyond $0.30$, thus the data has effectively roughly 30 significant dimensions.
After a number of initial experiments, we set $k=16$ clusters, as larger $k$ produced centroids similar to each other, while smaller $k$ mixed classes that are distinct for $k=16$. 
Figure~\ref{fig:clusters-inst} shows clusters' centroids, i.e., the average distribution from all the tasks assigned to a single cluster by the k-means algorithm.

First, although number of tasks in a cluster varies considerably (from 1.4\% to 17.6\% of all long tasks), no cluster strictly dominates the data, as the largest cluster groups less than 1/5th of all the tasks. 
Second, the centroids vary significantly.
Some of the centroids correspond to simple distributions, e.g., tasks that almost always have 0 CPU usage (c), (f), (k); or exponential-like distributions (b) and (p); while others correspond to mixed distributions (h), (j), (m).
Both observations demonstrate that no single probability distribution can describe all tasks in the trace.
Consequently, this data set does not satisfy the assumption that tasks follow a single distribution.

\begin{figure}[!tb]
\centering
  \subfloat[$N=10$]{{\includegraphics[width=.33\columnwidth]{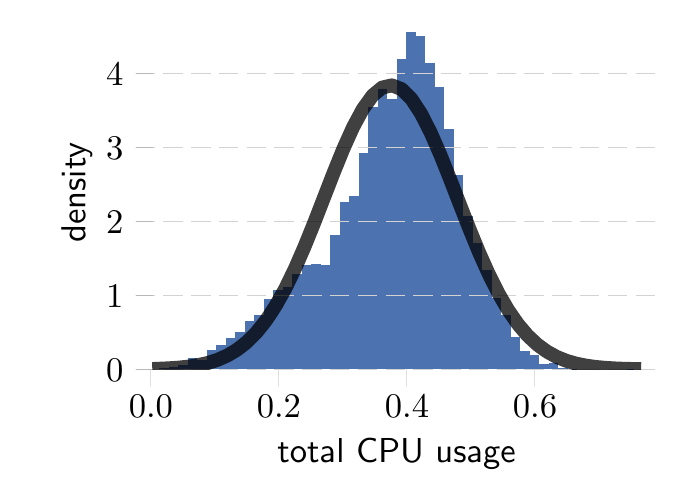}}}%
  \subfloat[$N=50$]{{\includegraphics[width=.33\columnwidth]{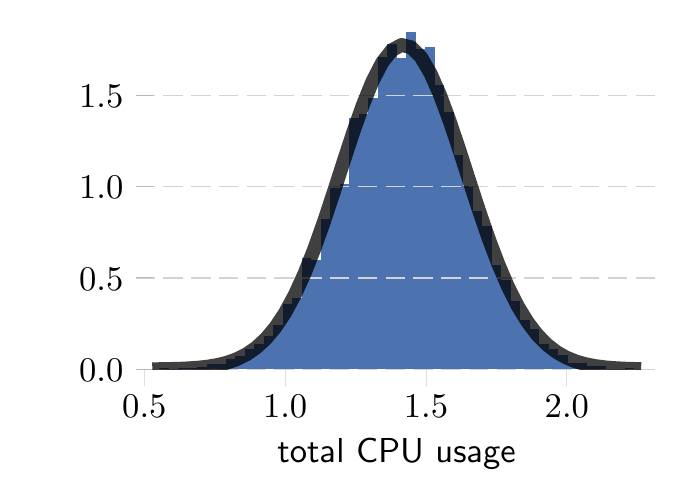}}}%
  \subfloat[$N=500$]{{\includegraphics[width=.33\columnwidth]{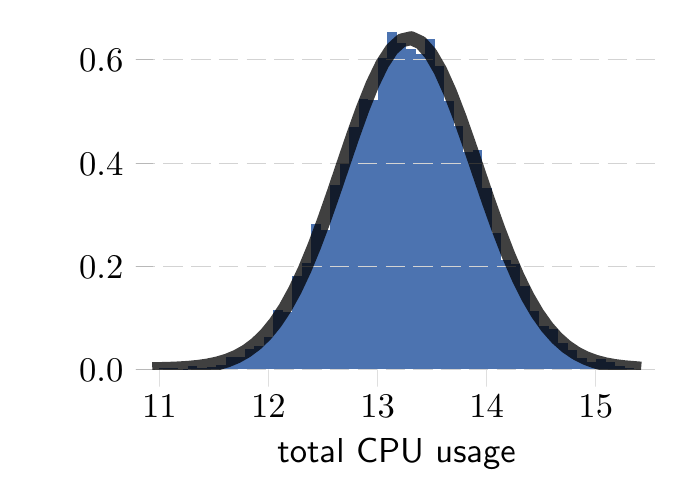}}}%
  \caption{Example empirical distributions of the total instantaneous CPU usage summed over randomly-chosen samples of $N=10$, $N=50$ and $N=500$ tasks. The black line denotes a fitted Gaussian distribution. Note the different ranges of X and Y axes.}
  \label{fig:cumul-instant-samples}
\end{figure}

\subsection{Characterizing the total usage}\label{sec:total-load-stats}

The previous section demonstrated that individual tasks' usage distributions are varied.
However, the scheduler is more concerned with the \emph{total} CPU demand of the tasks colocated on a single physical machine.
The central limit theorem states that the sum of random variables tends towards the Gaussian distribution (under some conditions on the variables, such as Lyapunov’s condition).
The question is whether the tasks in the Google trace are small enough so that, once the Gaussian approximation becomes valid, their total usage is still below the capacity of the machine.


To test this hypothesis, from our set of $100\,000$ tasks (Section~\ref{sec:data-preprocessing}), we take random samples $S_j$ of $N=\{10, 20, \dots, 500\}$ tasks (repeating the sampling $10\,000$ times for each size).
For each sample, we calculate the resulting empirical distribution based on $10\,000$ realizations of the instantaneous CPU usage.
Additionally, for each sample $S_j$, we fit a Gaussian distribution $N(\mu_j, \sigma_j)$.
We use standard statistics over the tasks $i \in S_j$ to estimate parameters:
$\mu_j = \sum_{i \in S_j} \mu_i$ and $\sigma_j = ( \sum_{i \in S_j} \sigma_i^2 )^{\frac{1}{2}}$,
where $\mu_i$ is the mean usage of task $i$, $\mu_i = \bar{x_i}$, and $\sigma_i$ its standard deviation.
Figure~\ref{fig:cumul-instant-samples} shows empirical distribution for three randomly-chosen samples and the fitted Gaussian distributions.

As the empirical distributions resemble the Gaussian distribution, we used the Anderson-Darling (A-D) test to check whether the resulting cumulative distribution is \emph{not} Gaussian (the A-D test assumes as the null hypothesis that the data is normally distributed with unknown mean and standard deviation).
Table~\ref{tab:h0-rejection-rates} shows aggregated results, i.e., 
fraction of samples of a given size $N$ for which the A-D test \emph{rejects} the null hypothesis at significance level of $5\%$.
A-D rejection rates for smaller samples (10-100 tasks) are high, thus the distributions are not Gaussian.
For instance assume that $N=50$ tasks are collocated on a single machine; if they are chosen randomly, in 78\% of cases the resulting distributions of total instantaneous CPU usage is not Gaussian according to the A-D test.
On the other hand, for 500 tasks, although A-D rejection rate is roughly 9\%, the mean cumulative usage is 13.5, i.e., $13.5$ times larger than the capacity of the largest machine in the cluster from which the trace was gathered.

\begin{table}[!tb]
  \centering
\begin{tabu} to \columnwidth {  X[l] | X[r] | X[r] | X[r] | X[r] | X[r] | X[r] | X[r] }
N & \textbf{10} & \textbf{20} & \textbf{30} & \textbf{50} & \textbf{100} & \textbf{250} & \textbf{500}\\
    \hline
    AD & 0.991 & 0.965 & 0.923 & 0.784 & 0.493 & 0.183 & 0.087\\
    $\bar{\mu_j}$ & 0.27 & 0.55 & 0.82 & 1.37 & 2.74 & 6.85 & 13.69 \\
\end{tabu}
  \caption{H0 rejection rates (middle row) for the normality of the total instantaneous CPU usage by the number of tasks in a sample. Anderson-Darling test at significance level of 5\%. The bottom row shows mean (over all samples) $\mu_j$, the estimated mean of the CPU usage for the given number of tasks.}
  \label{tab:h0-rejection-rates}
\end{table}

\begin{figure}
\centering
\subfloat[10 tasks in a bin]{{\includegraphics[width=0.5\columnwidth]{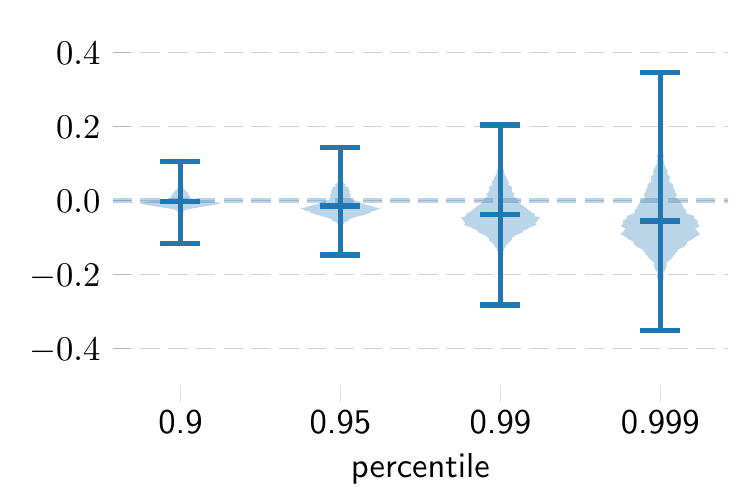}}}%
\subfloat[20 tasks in a bin]{{\includegraphics[width=0.5\columnwidth]{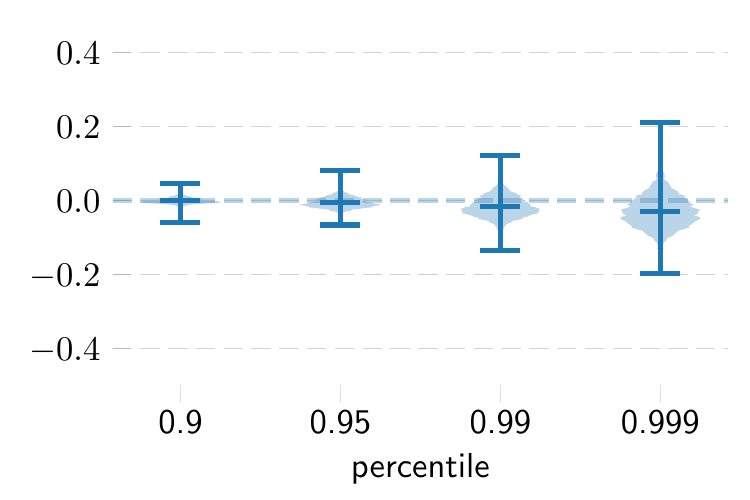}}}%
\\
\subfloat[30 tasks in a bin]{{\includegraphics[width=0.5\columnwidth]{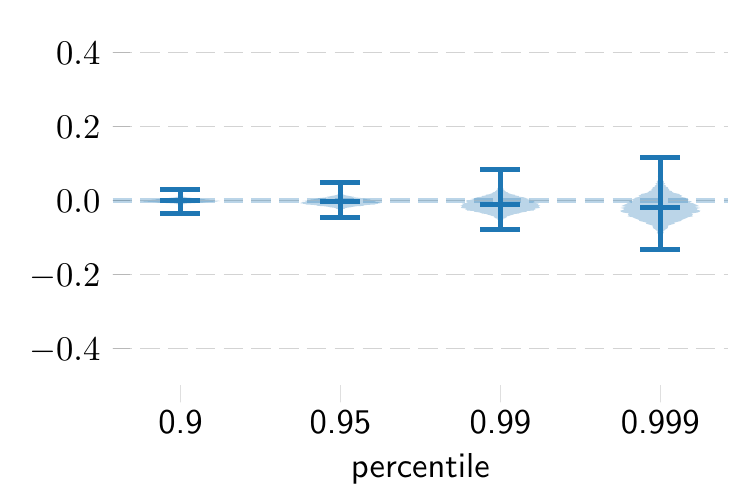}}}%
\subfloat[50 tasks in a bin]{{\includegraphics[width=0.5\columnwidth]{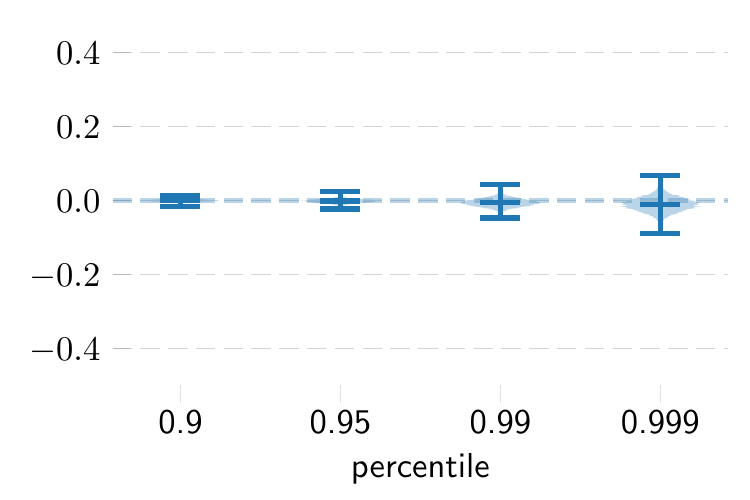}}}
\caption{Relative differences $(F^{-1}_{\mu_j, \sigma_j}(k) - P_k)/P_k$ between the k=95th, k=99th and the k=99.9th percentiles of the inverse CDF $F^{-1}$ of the normal distribution and the corresponding values of the empirical distribution $P_k$. Violin plots with outlines showing a (slightly smoothed) histogram and whiskers---the distributions of the differences. Each violin shows a statistics over 10 000 independent samples.
}
  \label{fig:percentiles-errors}
\end{figure}

However, to solve the packing problem, we need a weaker hypothesis. Rather than trying to predict the whole distribution, we only need the probability $\varrho$ of exceeding the machine capacity $c$; this probability corresponds to the value of the survival function ($1-CDF$) in a specific point $c$.
As our main positive result we show that it is possible to predict the value of the empirical survival function with a Gaussian usage estimation.
The following analysis does not yet take into account the machine capacity $c$; here we are generating a random sample $S_j$ of 10 to 50 tasks and analyze their total usage.
In the next section we show an algorithm that takes into account the capacity.

As $\varrho$ corresponds to the requested SLO, typically its values are small, e.g., 0.1, 0.01 or 0.001; these values correspond to the $k=90$th, $k=99$th or $k=99.9$th percentiles of the distribution. The question is thus how robust is the prediction of such a high percentile.

To measure the robustness, we compute the estimated usage at the $k$th percentile as 
the value of the inverse distribution function $F^{-1}_{\mu_j, \sigma_j}$ of the fitted Gaussian distribution $N(\mu_j, \sigma_j)$ at the requested percentile $k$.
We compare $F^{-1}_{\mu_j, \sigma_j}(k)$ with $P_k$, the $k$-th percentile of the empirical distribution.
Figure~\ref{fig:percentiles-errors} shows $(F^{-1}_{\mu_j, \sigma_j}(k) - P_k)/P_k$,
i.e., the relative difference between the Gaussian-based estimation and the empirical percentile.
We see that, first, medians are close to 0, which means that the Gaussian-based estimation of the total usage is generally accurate.
Second, the Gaussian-based estimation underestimates rare events, i.e., it underestimates the usage for high percentiles.
Third, if there are more tasks, the variance of the difference is smaller.


\section{Stochastic Bin Packing with Gaussian Percentile Approximation (GPA)}\label{sec:gpa-algorithm}

The main positive result from the previous section is that a Gaussian estimation estimates  values of high percentiles of the total instantaneous CPU usage. In this section, we formalize this observation into GPA, a fit test that uses the central limit theorem to drive a stochastic bin packing algorithm. GPA stems from statistical multiplexing, a widely-used approach in telecommunications, in which individual transmissions, each with varying bandwidth requirements, are to be packed onto a communication channel of a fixed capacity (although our models, following~\cite{kleinberg_allocating_2000,goel1999stochastic}, do not consider packet buffering, making the models considerably easier to tackle). A related test, although assuming that the packed items all have Gaussian distributions, was proposed in~\cite{wang_consolidating_2011} for multiplexing VM bandwidth demands.

A standard bin-packing algorithm (such as First Fit, Best Fit, etc.) packs items $\{ x_i \}$ to bins $S_j$ sequentially. For instance, the First Fit algorithm, for each item $x_k$, finds the minimal bin index $j$, such that $x_k$ fits in the bin $S_j$. The fitting criterion is simply that the sum of the sizes of items $S_j$ already packed in $j$ and the current item $x_k$ is smaller than the bin capacity $c$, $x_k + \sum_{x_i \in S_j} x_i \leq c$.

\SetInd{0.3em}{0.3em}
\begin{algorithm}[!tb]
  \footnotesize
  \SetKwInput{KwNotation}{Notation}
  \SetKwFunction{Mean}{Mean}
  \SetKwFunction{Std}{Std}
  \SetKwFunction{CDF}{F}
  \SetKwFunction{Cap}{Cap}
  \SetKwFunction{EmptyBin}{EmptyBin}
  \SetKwFunction{FitBin}{FitBin}
  \SetKwFunction{FindBin}{FindBin}
  \SetKwBlock{Block}
  \SetAlCapFnt{\footnotesize}
  \KwNotation{\\
     \CDF{$x|\mu, \sigma^2$}---the value of the CDF of the Gaussian distribution $\mathcal{N}(\mu, \sigma^2)$ in a point $x$\\
  }
  
  \vspace{1em}

  \FitBin{$k$, $j$, $\rho$}
  \Block{
    $\mu_j' = \mu_k + \sum_{i \in S_j} \mu_{i}$\;
    $\sigma_j' = ( \sigma_k^2 + \sum_{i \in S_j} \sigma_i^2 )^{\frac{1}{2}}$\;
    return $\rho - \big( 1 - \CDF{$c|\mu_j', (\sigma_j')^2$} \big)$\;

  }

  \vspace{1em}

  \FindBin{$k$, $\rho$}
  \Block{
    \For{$j$ in $1..m$}{
      \If{$\FitBin{$k$, $j$, $\rho$} \ge 0$}{
        return $j$\;
      }
    }
    {
      $m \leftarrow m+1$\;
      return $m$\;
    }
  }
  \caption{GPA algorithm: find the first machine $j$ to which task $t$ fits.}
  \label{alg:gpa-findbin}
\end{algorithm}

Our method, the Gaussian Percentile Approximation (GPA,~\ref{alg:gpa-findbin}) replaces the  fitting criterion with an analysis of the estimated Gaussian distribution.
For each open (i.e., with at least one task) machine $j$, we store the current estimation of the mean $\mu_j$ and of the standard deviation $\sigma_j$.
We use standard statistics over the tasks $i \in S_j$ to estimate these values: $\mu_j = \sum_{i \in S_j} \mu_i$ and $\sigma_j = ( \sum_{i \in S_j} \sigma_i^2 )^{\frac{1}{2}}$.
When deciding whether to add a task $k$ to a machine $j$, we recompute the statistics taking into account task $k$'s mean $\mu_k$ and standard deviation $\sigma_k$: $\mu_j' = \mu_j + \mu_k$; $\sigma_j' = ( \sigma_k^2 + \sum_{i \in S_j} \sigma_i^2 )^{\frac{1}{2}}$. 
The task $k$ fits in the machine $j$ if and only if the probability that the total usage exceeds $c$ is smaller than $\varrho$.
We use the CDF of the Normal distribution $\CDF_{\mu_j', \sigma_j'}$ to estimate this probability, i.e., a task fits in the bin if and only if $\CDF_{\mu_j', \sigma_j'}(c) \geq 1 - \varrho$.

Algorithm~\ref{alg:gpa-findbin} shows First Fit with GPA. Best Fit can be extended analogously: instead of returning the first bin for which \texttt{FitBin $\ge$ 0}, Best Fit chooses a bin that results in minimal among positive \texttt{FitBin} results.

As both First Fit and Best Fit are greedy, usually the last open machine ends up being underutilized.
Thus, after the packing algorithm finishes, 
to decrease the probability of overload in the other bins, we rebalance the loads of the machines  by a simple heuristics. Following the round robin strategy, we choose a machine from $\{1, \dots, m-1\}$, i.e., all but the last machine. Then we try to migrate its first task to the last machine $m$: such migration fails if the task does not fit into $m$. The algorithm continues until $max\_failures$ failed attempts (we used $max\_failures=5$ in our experiments). Note that many more advanced strategies are possible. Any such rebalancing makes the algorithm not on-line as it reconsiders the previously taken decisions  (in contrast to First Fit or Best Fit, which are fully on-line).

  
  

\section{Validation of GPA through Simulation Experiments}\label{sec:experiments}
The goal of our experiments is to check whether GPA provides empirical QoS similar to the requested SLO while using a small number of machines.

\subsection{Method}\label{sec:sim-method}
Our evaluation relies on Monte Carlo methods.
As input data, we used our random sample of $N=100\,000$ tasks, each having $R=10\,000$ realizations of the instantaneous and the 5-minute average loads generated from empirical distributions (see Section~\ref{sec:data-preprocessing}).
To observe algorithms' average case behavior we further sub-sample these $N=100\,000$ tasks into 50 \emph{instances} each of $N'=1\,000$ tasks.
A single instance can be thus compactly described by two matrices $x[i][t]$ and $y[i][t]$, where $x$ denotes the instantaneous and $y$ the 5-minute average usage; $i \in \{1, \dots, N'\}$ is the index of the task and $t \in \{ 1, \dots, R \}$ is the index of the realization.

Many existing theoretical approaches to bin-packing implicitly assume clairvoyance, i.e., the sizes of the items are known in advance.
We test the impact of this assumption by partitioning the matrices' columns into the observation and the evaluation sets.
The bin-packing decision is based only on the data from the observation set $O$, while to evaluate the quality of the packing we use the data from the evaluation set $E$ ($O$ and $E$ partition $R$, i.e., $O \cap E =\emptyset$ except in the fully clairvoyant scenario, in which $O=E=R$).
For instance, we might assume we are able to observe each tasks' 100 instantaneous usage samples before deciding where to pack it: this corresponds to the observation set $O = \{ 1, \dots, 100 \}$, i.e., $x[i][1 \dots 100]$, and the evaluation $E = \{ 101 \dots 10\,000 \}$, set of $x[i][101 \dots 10\,000]$.
In this case our algorithms will compute statistics based on $x[i][1 \dots 100]$ (e.g., the Gaussian Percentile Approximation will compute the mean $\mu_i$ as $\mu_i = \frac{1}{100} \sum_{t=1}^{100} x[i][t]$).
As we argued in Section~\ref{sec:definition}, scenarios with limited clairvoyance simulate varying quality of prediction of the resource manager.
An observation set equal to the evaluation set corresponds to clairvoyance, or the perfect estimations of the usage; smaller observation sets correspond to worse estimations.

We execute GPA with four different target SLO levels, $\varrho=0.10$ corresponding to the SLO of $90\%$; $\varrho=0.05$ (SLO of $95\%$); $\varrho=0.01$ (SLO $99\%$); and $\varrho=0.001$ (SLO $99.9\%$).

We compare GPA with the following estimation methods:
\renewcommand\descriptionlabel[1]{$\bullet$ \textbf{#1}}
\begin{description}
\item[Cantelli] (proposed for the data center resource management in~\cite{hwang_hierarchical_2016}): items with sizes equal to the mean increased by $b$ times the standard deviation. According to Cantelli's inequality, for any distribution $Pr[ X > \mu + b\sigma ] < \frac{1}{1+b^2}$. Thus $b=4.4$ ensures SLO of $95\%$. If $X$ has a Normal distribution (which is not the case for this data, Figure~\ref{fig:clusters-inst}), the multiplier can be decreased to $b=1.7$ with keeping SLOs at the same level~\cite{hwang_hierarchical_2016}. In initial experiments, we found those values to be too conservative and decided to also consider an arbitrary multiplier $b=1.0$.
\item[av]: Items with sizes proportional to the \emph{mean} $\mu_i$ from the observation period. The mean is multiplied by a factor $f \in \{ 1.0, 1.25, 2.0 \}$. Factors larger than $1.0$ leave some margin when items' realized size is larger than its mean. 
\item[perc]: Items with sizes equal to a certain percentile from the observation period. We use the $\{50, 70, 90, 95, 99, 100\}$th percentile. The \emph{maximum} (100th percentile) corresponds to a conservative policy that packs items by their true maximums---this policy in fully clairvoyant scenario is essentially packing tasks by their observed maximal CPU consumption.
Lower percentiles correspond to increasingly aggressive policies.
\end{description}

All these methods use either First Fit or Best Fit as the bin packing algorithm. We analyze the differences between the two in Section~\ref{sec:exp-comp-algorithms}. All use rebalancing; we analyze its impact in Section~\ref{sec:exp-repackaging}.

We use two metrics to evaluate the quality of the algorithm. The first metric is the number of used machines (opened bins).
This metric directly corresponds to the resource owner's goal---using as few resources as possible.
Different instances might have vastly different usage, resulting in different number of required machines.
Thus, for a meaningful comparison,
for each instance we report a \emph{normalized} number of machines $m$.
We compute $m$ as $m=m_{\text{abs}} / m_{\text{norm}}$, i.e., $m_{\text{abs}}$, the number of machines used by the packing algorithm, divided by a normalization value $m_{\text{norm}}$.
The normalization value computes the total average CPU requirements in the instance, and then divides it by machine capacity, $m_{\text{norm}} = \lceil \frac{1}{c} \sum_i \bar{x_i} \rceil$.

The second metric is the measured frequency $q$ of exceeding the machine capacity $c$: the higher the $q$, the more often the machine is overloaded. $(1-q)$ corresponds to the observed (empirical) QoS.
We compute $q$ by counting $q(j)$: independently for each machine $j$, how many of $E$ realizations of the total instantaneous usage resulted in total machine usage higher than $c$: $q(j) = \sum_{t \in E} \langle ( \sum_{i \in S_j}x[i][t] ) > c \rangle $, where $\langle pred \rangle$ returns 1 if the predicate $pred$ is true. We then average these values over all $m$ machines and the complete evaluation period $E$, $q =\frac{1}{m|E|} \sum_{j=1}^m q(j)$.

The base case for the experiments is the full clairvoyance (i.e., observation set equal to the evaluation set, $O=E=R$); machine capacity $c=1$; estimation based on instantaneous (\emph{inst}) data. The following sections test the impact of these assumptions.

\subsection{Comparison between algorithms}\label{sec:exp-comp-algorithms}
Both FirstFit and the BestFit algorithms lead to similar outcomes.
The number of machines used $m_{\text{abs}}$ differed by at most 1. The mean values for $q$ differed by less than 1\%.
Consequently, to improve presentation, we report data just for BestFit.

The tasks in the trace have small CPU requirements, as already shown in Figure~\ref{fig:cumul-instant-samples}, which reported for 500 tasks the mean total requirement of $13.5$. \emph{av 1}, packing tasks by their mean requirements, confirms this result, packing 1000-task instances into, on the average, $\bar{m}_{\text{abs}}=28.0$ machines.

\begin{figure}[!tb]
  \includegraphics[trim={.7cm .5cm 1cm 0},clip,width=\columnwidth]{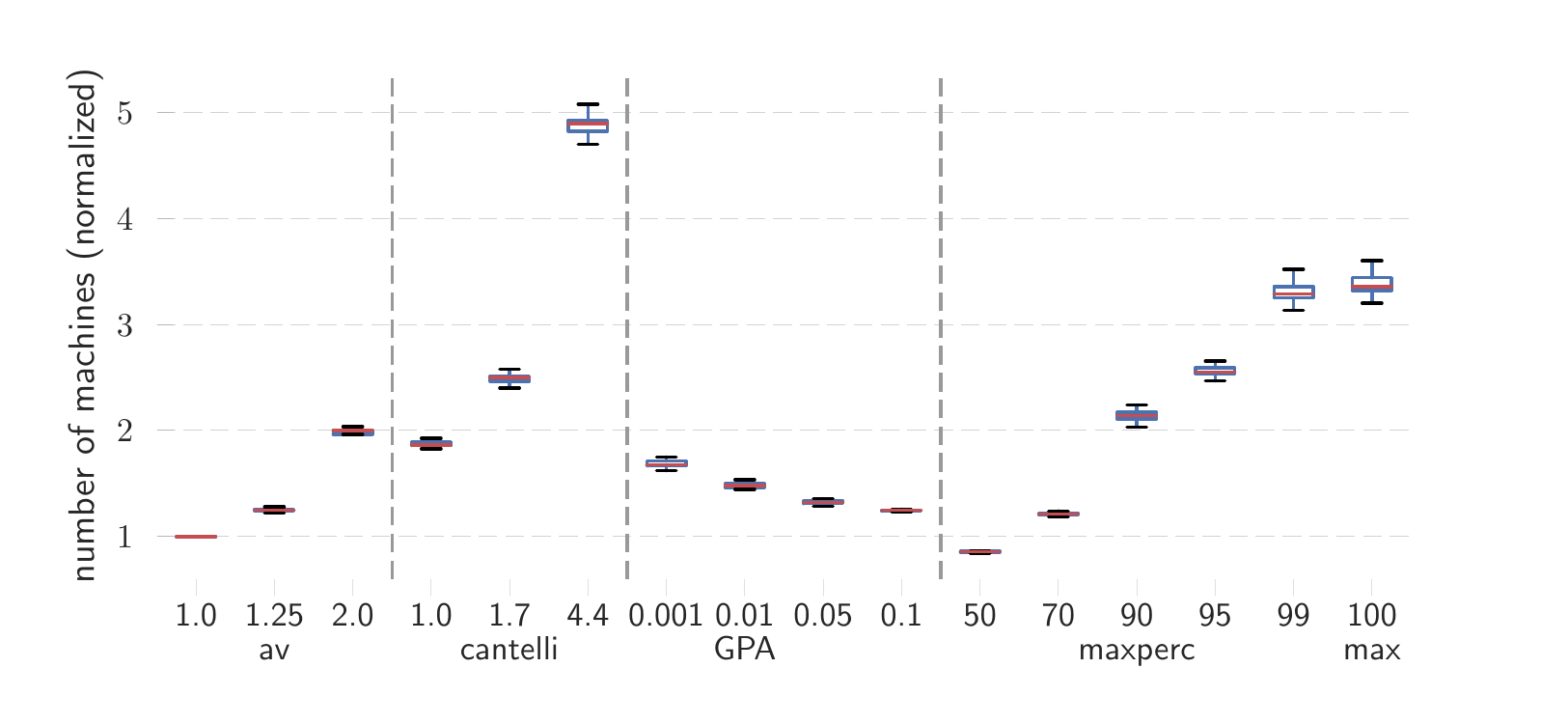}
  \caption{Number of machines (normalized to the lower bound) by different estimation algorithms. Clairvoyance, $c=1$. Here and in the remaining boxplots, the statistics for each box are computed over 50 instances.}\label{fig:exp-clairvoyance-bins}
\end{figure}

The \emph{Cantelli} estimation is very conservative (for $b=1.0$, mean $\bar{q}=3\times10^{-4}$; for $b=1.7$, mean $\bar{q}=2\times10^{-6}$).
The \emph{max} estimation never exceeds capacity; and \emph{perc} with high percentiles is similarly conservative: 99th percentile leads to $\bar{q}=2\times10^{-8}$; the 95th to $\bar{q}=5\times10^{-6}$; and the 90th to $\bar{q}=7\times10^{-5}$.
The resulting packings use significantly more machines than \emph{av} and \emph{GPA} estimations (Figure~\ref{fig:exp-clairvoyance-bins}).

\begin{figure}[!tb]
  \includegraphics[trim={.3cm 0 1cm 0},clip,width=\columnwidth]{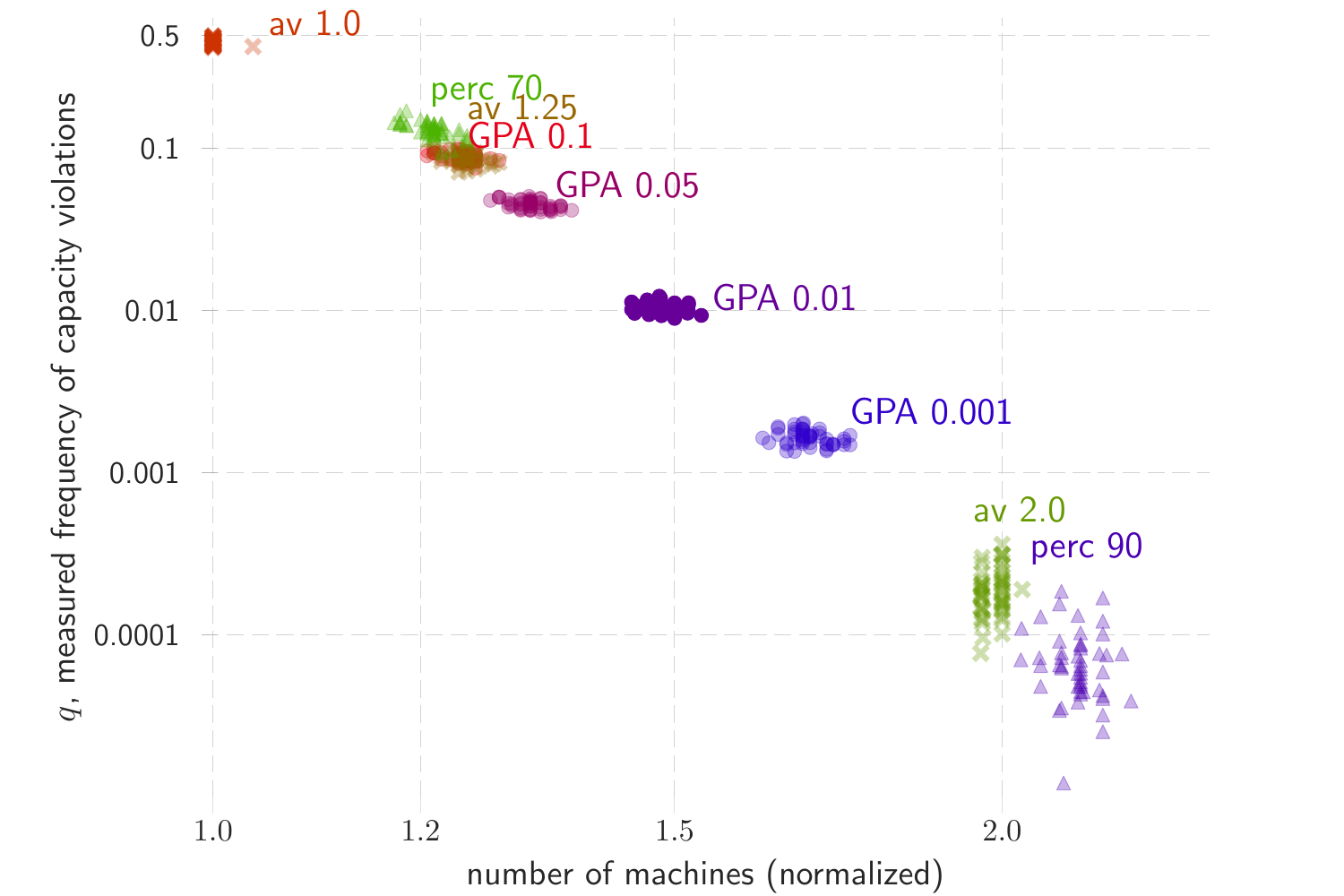}
  \caption{Comparison of the number of used machines (X axis, normalized to the lower bound) and the empirical frequency of capacity violations $q$ (Y axis) between \emph{GPA}, \emph{av} and \emph{perc}. Each dot represents a single instance. Clairvoyance, $c=1$}\label{fig:exp-gpa-avg-scatter}
\end{figure}

Figure~\ref{fig:exp-gpa-avg-scatter} shows the normalized number of machines $m$ and the empirical capacity violations $q$ for the remaining algorithms (\emph{GPA}, \emph{av} and \emph{perc}).
No estimation Pareto-dominates others---different methods result in different machine-QoS trade-offs.
The resulting $(m,q)$ can be roughly placed on a line, showing that $q$ decreases exponentially with an increase in the number of machines, $m$.
\emph{perc 70}, \emph{av 1.25} and \emph{GPA 0.1} result in comparable $m$-$q$; and, to somewhat smaller degree, \emph{perc 90}, \emph{av 2.0} and \emph{GPA 0.001}. Such similarities might suggest that to achieve a desired $q$, it is sufficient to use \emph{av} with an appropriate value of the multiplier. We test this claim in Section~\ref{sec:exp-varied-capacities}.


\begin{figure}[tbp]
  \includegraphics[width=\columnwidth]{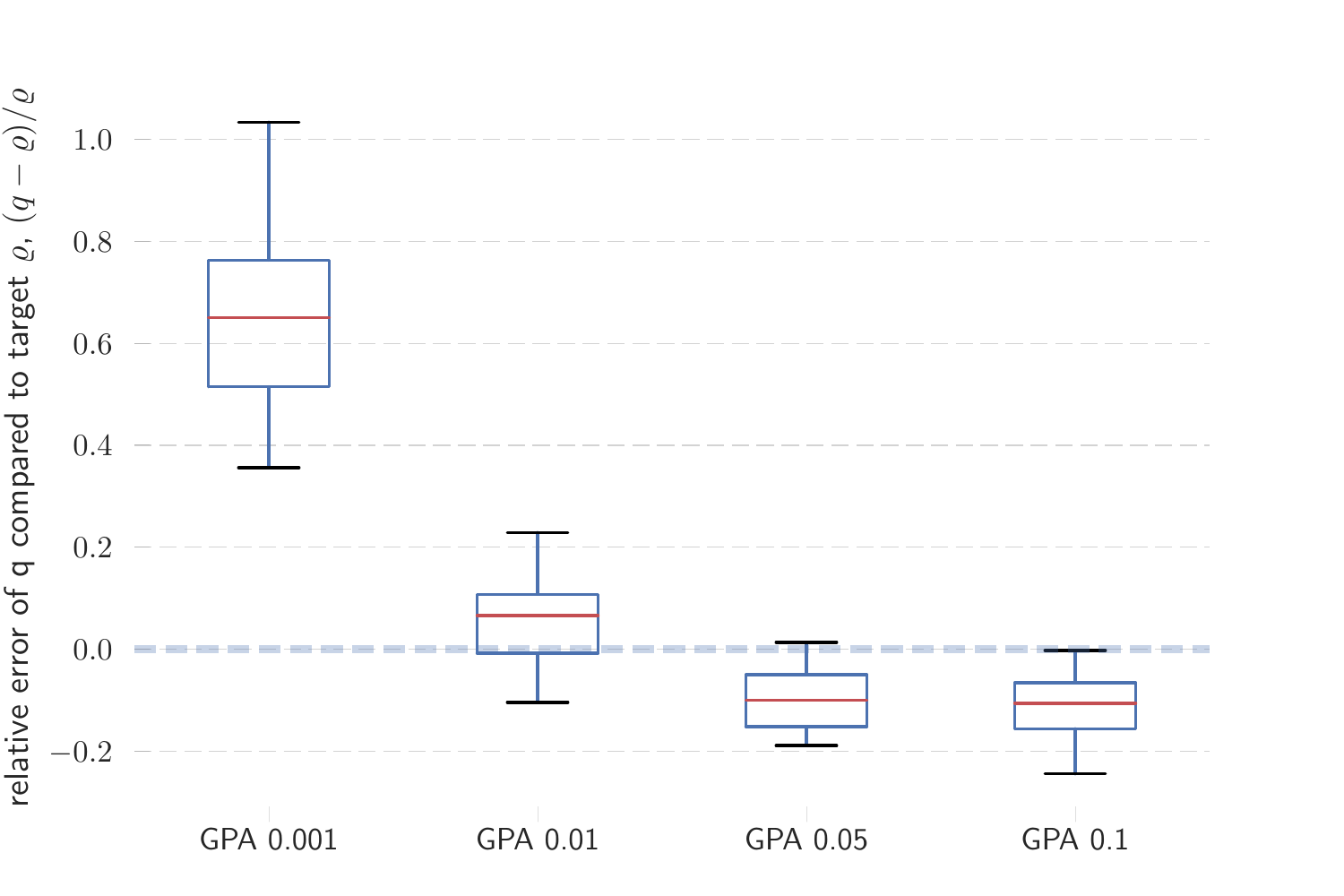}
  \caption{Relative error $(q-\varrho)/\varrho$ of \emph{GPA} for various requested SLO $\varrho$ values. Clairvoyance, $c=1$.}\label{fig:exp-gpa-relerror}
\end{figure}

Figure~\ref{fig:exp-gpa-relerror} analyses for GPA the relative error between the measured frequency of capacity violations $q$ and the requested SLO $\varrho$, i.e.: $(q-\varrho)/\varrho$. The largest relative error is for the smallest target $\varrho=10^{-3}$: GPA produces packings with the measured frequency of capacity violations of $q=1.6 \times 10^{-3}$, an increase of 60\%.
This result follows from the results on random samples (Figure~\ref{fig:percentiles-errors}): estimating the sum from the Gaussian distribution underestimates rare events.
As a consequence, for SLOs of 99th or 99.9th percentile, the $\varrho$ parameter of GPA should be adjusted to a smaller value.

\subsection{Observation of the 5-minute average usage}
\begin{figure}[tbp]
\centering
  \subfloat[\emph{av $2$}]{{\includegraphics[width=.25\columnwidth]{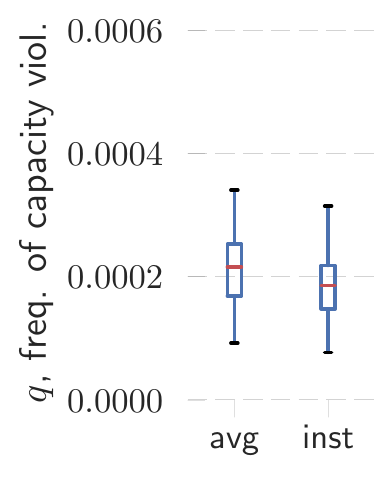}}}%
  \subfloat[\emph{av $1.25$}, ]{{\includegraphics[width=.25\columnwidth]{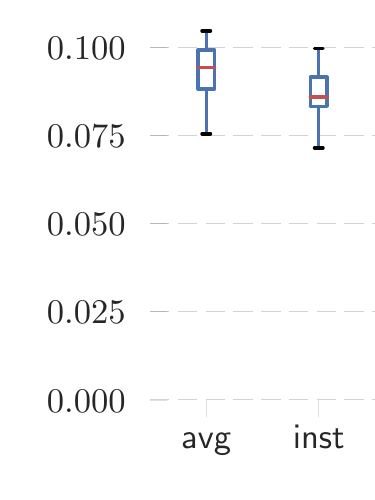}}}%
  \subfloat[\emph{Cantelli}, $1.0$]{{\includegraphics[width=.25\columnwidth]{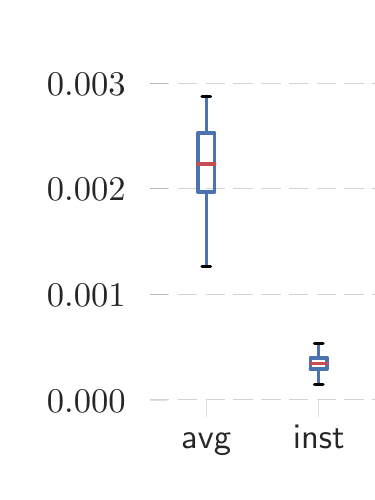}}}%
  \subfloat[\emph{Cantelli}, $1.7$]{{\includegraphics[width=.25\columnwidth]{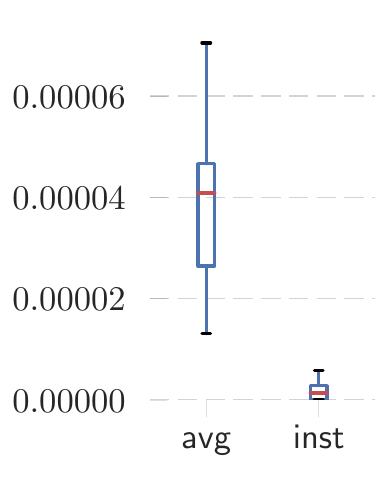}}}%
  \\
  \subfloat[\emph{GPA} $0.001$]{{\includegraphics[width=.25\columnwidth]{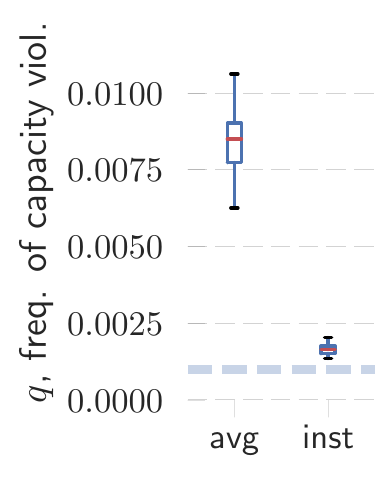}}}%
  \subfloat[\emph{GPA} $0.01$]{{\includegraphics[width=.25\columnwidth]{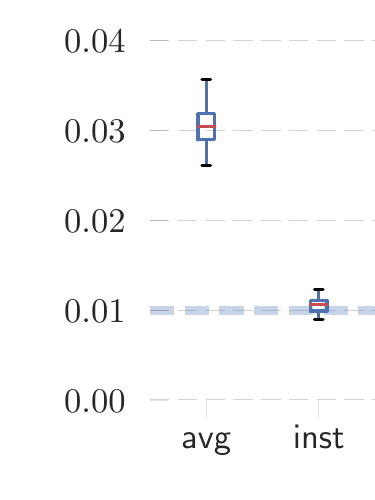}}}%
  \subfloat[\emph{GPA} $0.05$]{{\includegraphics[width=.25\columnwidth]{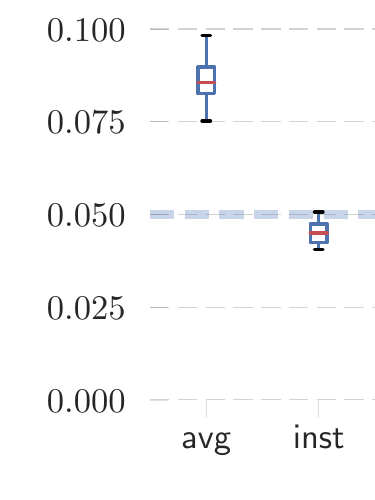}}}%
  \subfloat[\emph{GPA} $0.10$]{{\includegraphics[width=.25\columnwidth]{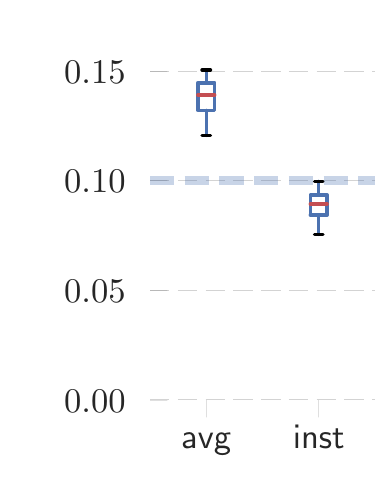}}}%
  \caption{Comparison of the measured frequency of capacity violations $q$ when each algorithm uses either the 5-minute averages (\emph{avg}), or 1-second instantaneous (\emph{inst}) data. Clairvoyance, $c=1$. For GPA, the target SLO is marked by a thicker line. Note that Y scales differ (see the discussion in Section~\ref{sec:exp-comp-algorithms} for comparison between these algorithms).}\label{fig:avg-obs}
\end{figure}

In this series of experiments we show that if algorithms use statistics over 5-minute averages  (the low-frequency data), the resulting packing has more capacity violations.
Estimations that use statistics of tasks' variability (such as the standard deviation in GPA and Cantelli) are more sensitive to less accurate avg data. This is not a surprise: as we demonstrated in Section~\ref{sec:data-analysis}, the averages report smaller variability than instantaneous usage.

Figure~\ref{fig:avg-obs} summarizes $q$ for various estimation methods. The figure does not show Cantelli with $b=4.4$, as on both datasets the mean $\bar{q}$ is 0.
Similarly, for \emph{max} (\emph{perc 100}) estimation, the mean $\bar{q}$ using 5-minute averages (\emph{avg}) is very small (albeit non-zero, in contrast to \emph{inst}): $3 \times 10^{-7}$ for $e=0.8$ and $5 \times 10^{-5}$ for bin size multiplier $e=1.0$.
High frequency \emph{inst} data significantly reduces the number of capacity violations for estimations that use the standard deviation. For Cantelli, the improvement is roughly 10 times; for GPA roughly 2-3 times.
In contrast, as expected, \emph{av} estimation has similar $q$ for both instantaneous and 5-minute average observations.

\subsection{Smaller and larger machines' capacities}\label{sec:exp-varied-capacities}

\begin{figure}[tbp]
\centering
  \subfloat[av 2, GPA 0.001, rebalancing]{{\includegraphics[width=.5\columnwidth]{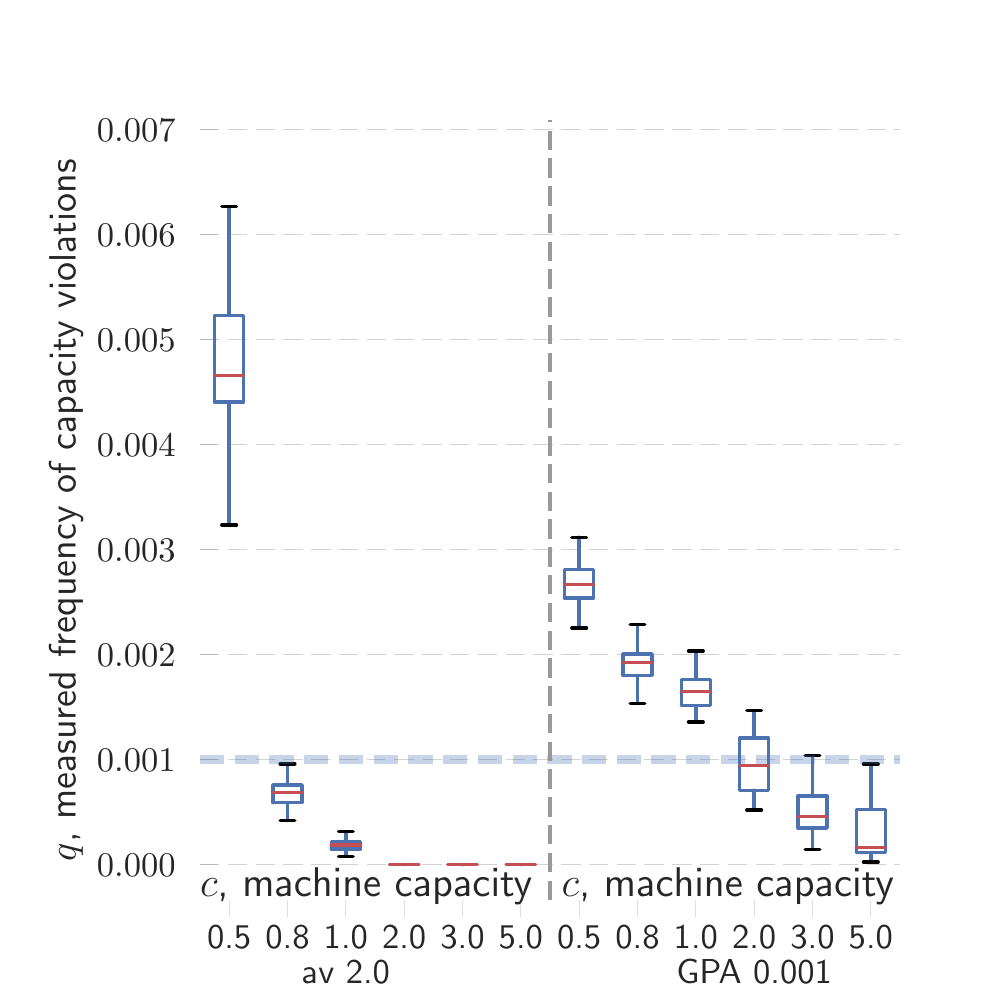}}}
  \subfloat[av 2, GPA 0.001, no rebalancing]{{\includegraphics[width=.5\columnwidth]{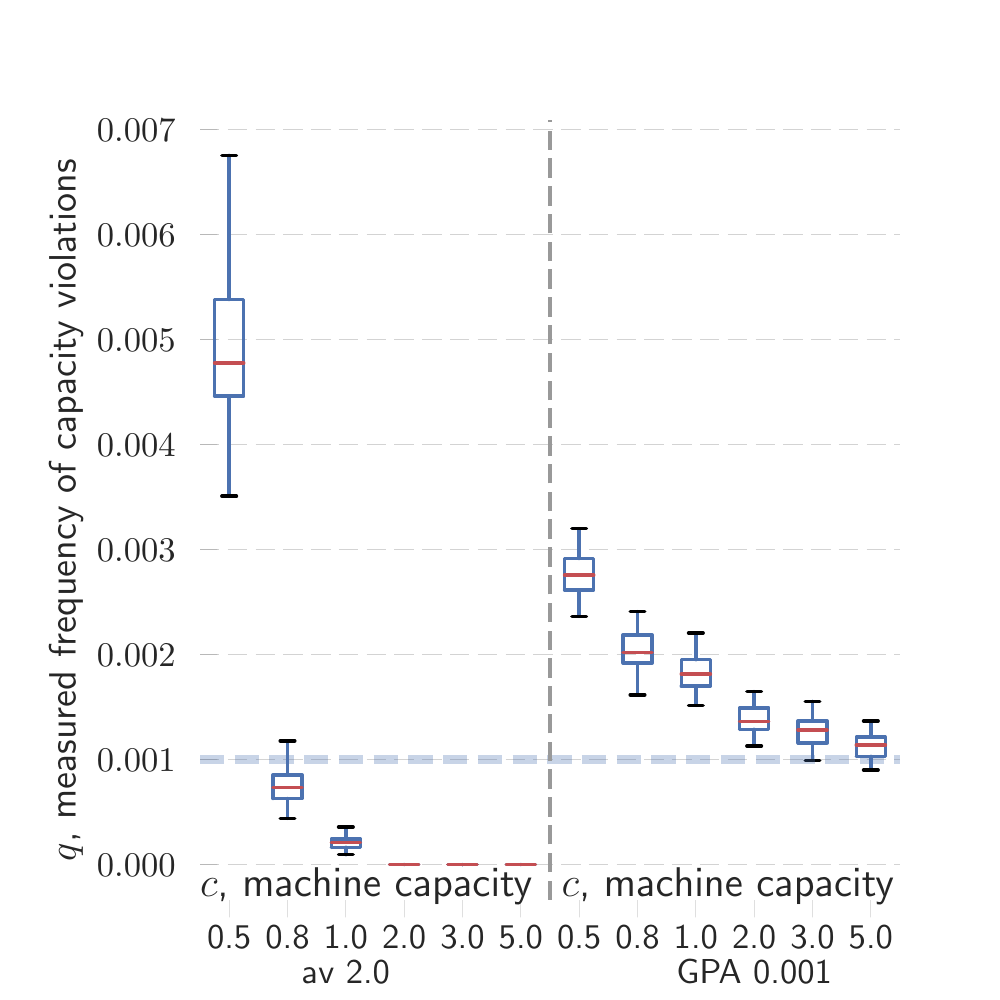}}}
  \\
  \subfloat[av 1.25, GPA 0.1, rebalancing]{{\includegraphics[width=.5\columnwidth]{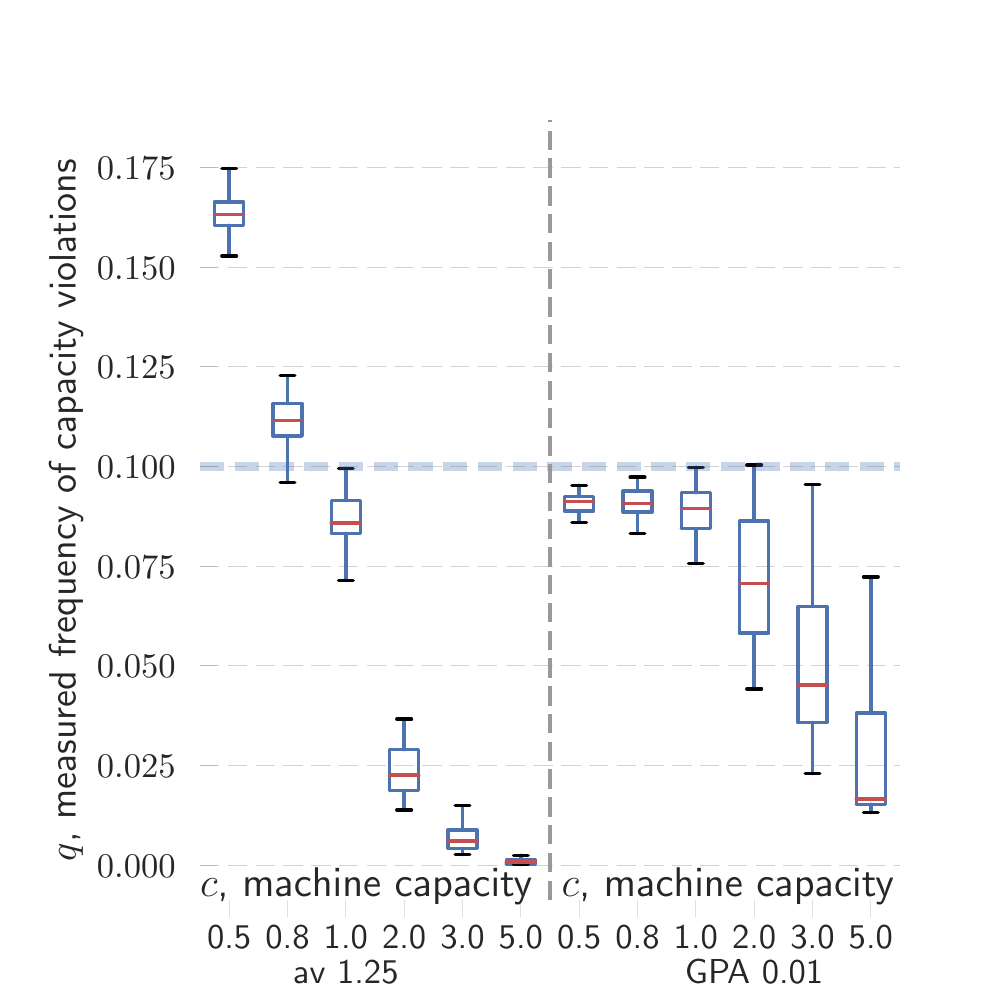}}}
  \subfloat[av 1.25, GPA 0.1,no rebalancing]{{\includegraphics[width=.5\columnwidth]{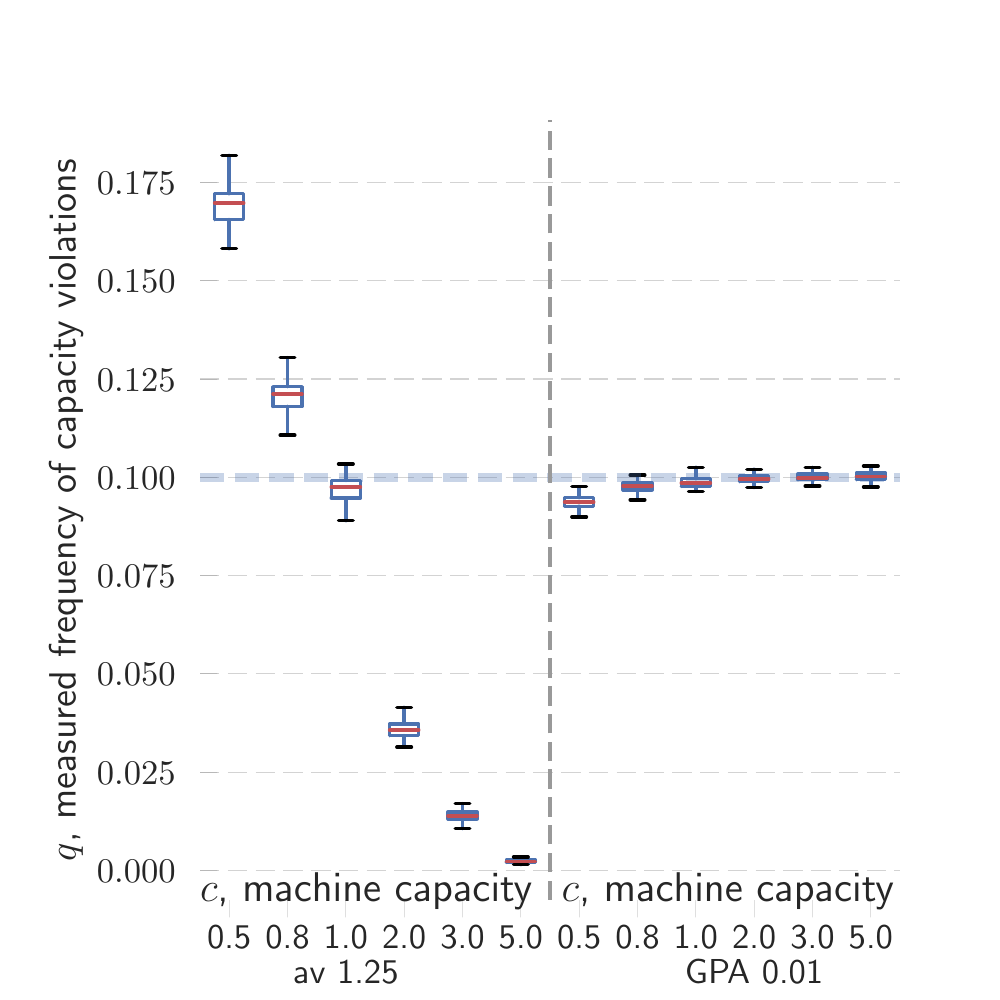}}}
  \caption{$q$ for \emph{av} and GPA by different machine capacities $c \in \{ 0.5, 0.8, 1.0, 2.0, 5.0 \}$.}\label{fig:smaller-capacities}
\end{figure}

By varying the bin capacity $c$, we are able to simulate different ratios of job requirements to machine capacity. (Note that for $c<1$, some of the tasks might not fit into any available machine having, e.g., the mean usage greater than $c$; however, as large tasks are rare, it was not the case for the 50 instances considered in the experiments).
As both the number of machines used and $q$ are normalized, we expect these values to be independent of $c$.
Figure~\ref{fig:smaller-capacities} compares \emph{av 2} to \emph{GPA 0.001}; and \emph{av 1.25} to \emph{GPA 0.1} as for capacity $c=1$ these pairs resulted in similar $(q,m)$ combinations (see Figure~\ref{fig:exp-gpa-avg-scatter}).

Overall, GPA results in similar $q$ for different machine capacities.
The differences in GPA results can be explained by two effects. When capacities are smaller, the effects of underestimating $q$ (observed in Figure~\ref{fig:exp-gpa-relerror}) are more significant. When capacities are larger, GPA results in less capacity violations than the requested thresholds. This is the impact of the last-opened bin which, with high probability, is underloaded; for larger capacities this last bin is able to absorb more tasks during the rebalancing phase. Figures~\ref{fig:smaller-capacities}~(b) and~(d), where we measure $q$ for algorithms without rebalancing, confirms this explanation.

In contrast, for \emph{av}, $q$ differs significantly when capacities change.
This result demonstrates that using fixed thresholds for \emph{av} estimation on heterogeneous resources results in unpredictable frequency of capacity violations: thresholds have to be calibrated by trial and error, and a threshold achieving a certain QoS for a certain machine capacity results in a different QoS for a different machine capacity.

\subsection{Clairvoyance}\label{sec:exp-clairvoyance}
\begin{figure}[tbp]
\centering
\subfloat[GPA 0.01]{{\includegraphics[width=.5\columnwidth]{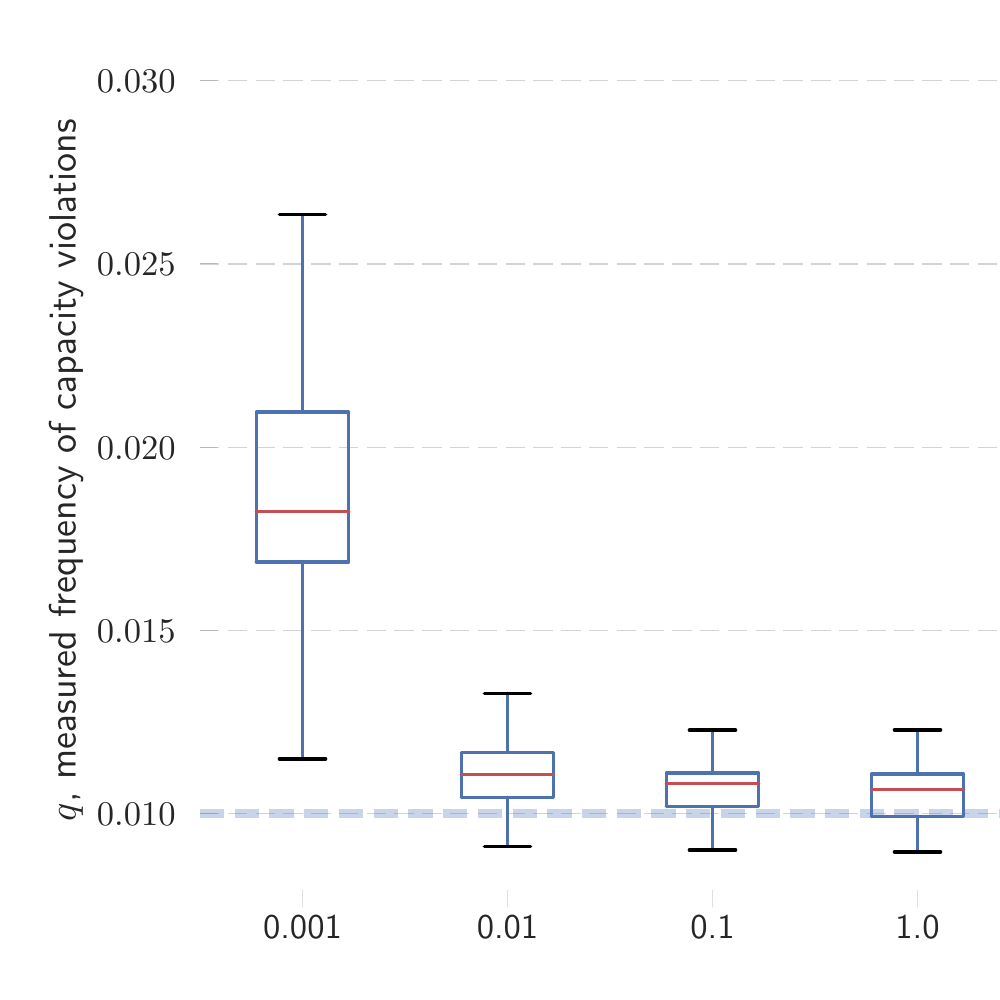}}}
\subfloat[av 1.25]{{\includegraphics[width=.5\columnwidth]{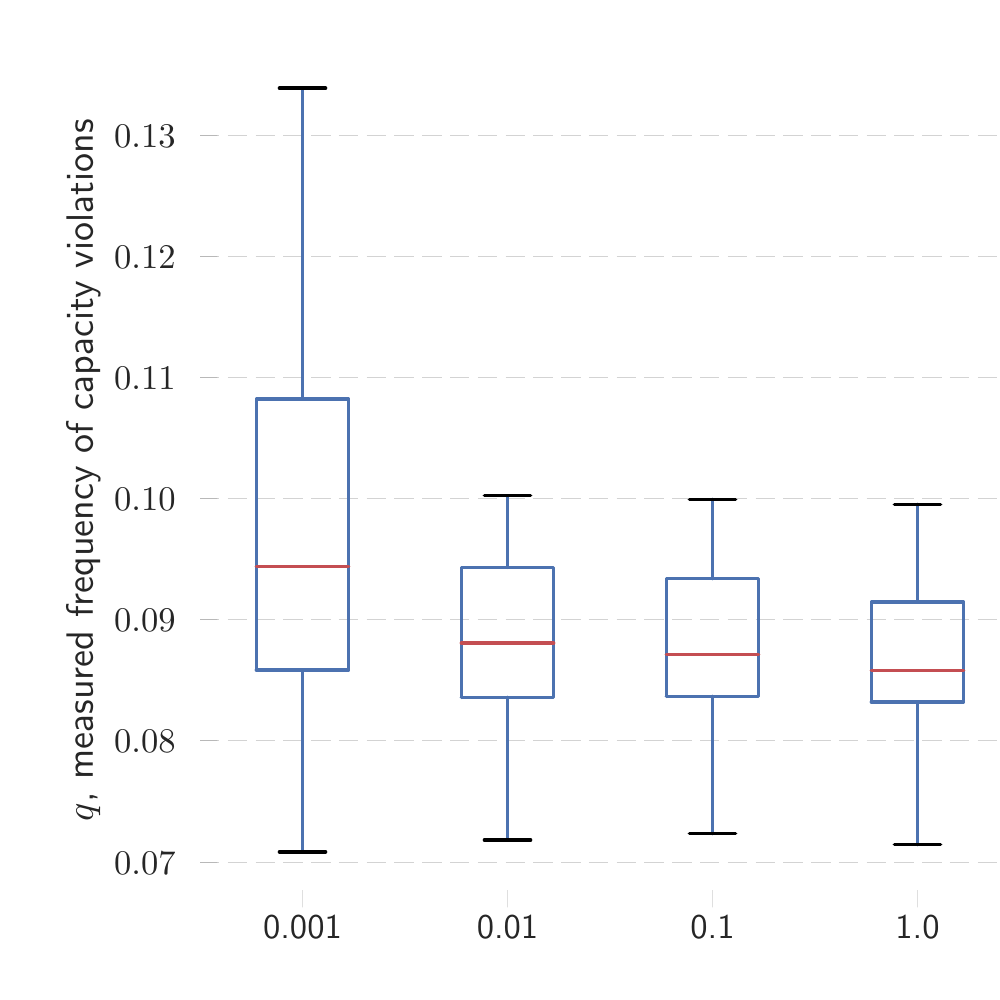}}}%
\\
\subfloat[GPA 0.001]{{\includegraphics[width=.5\columnwidth]{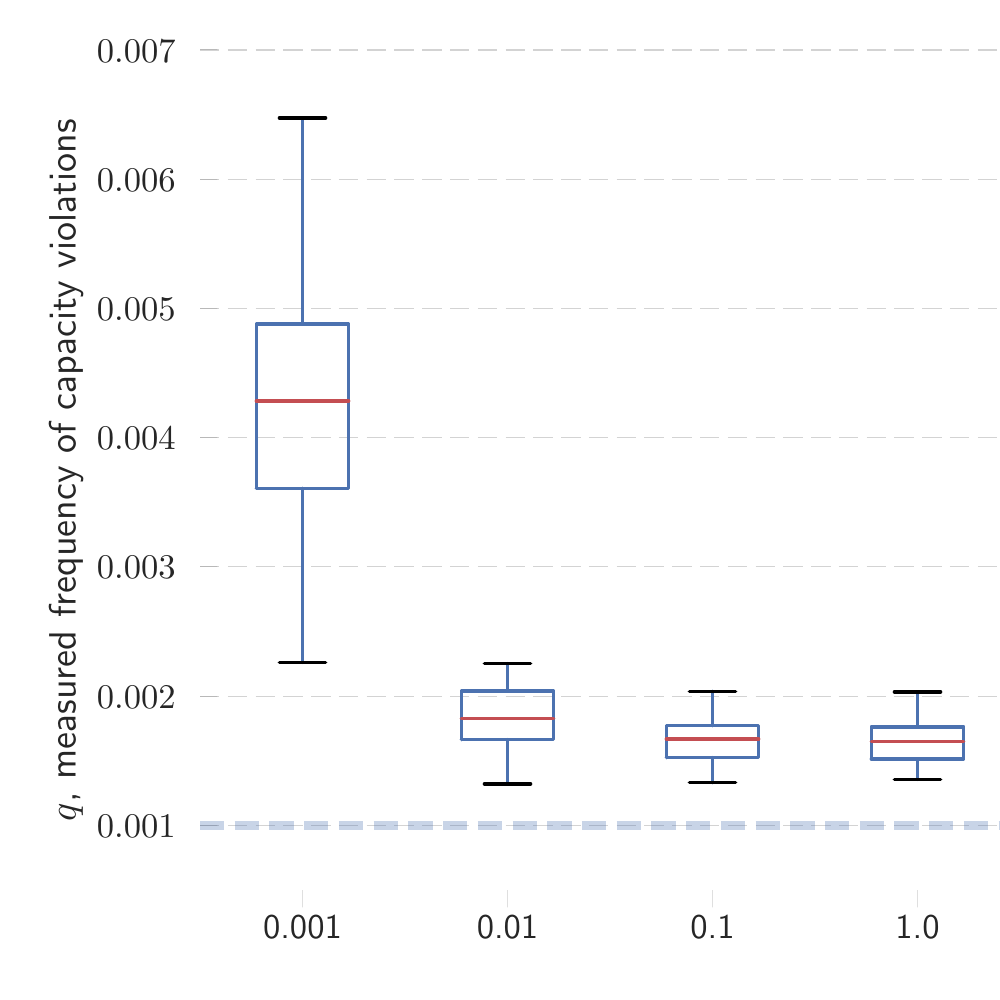}}}
\subfloat[av 2.0]{{\includegraphics[width=.5\columnwidth]{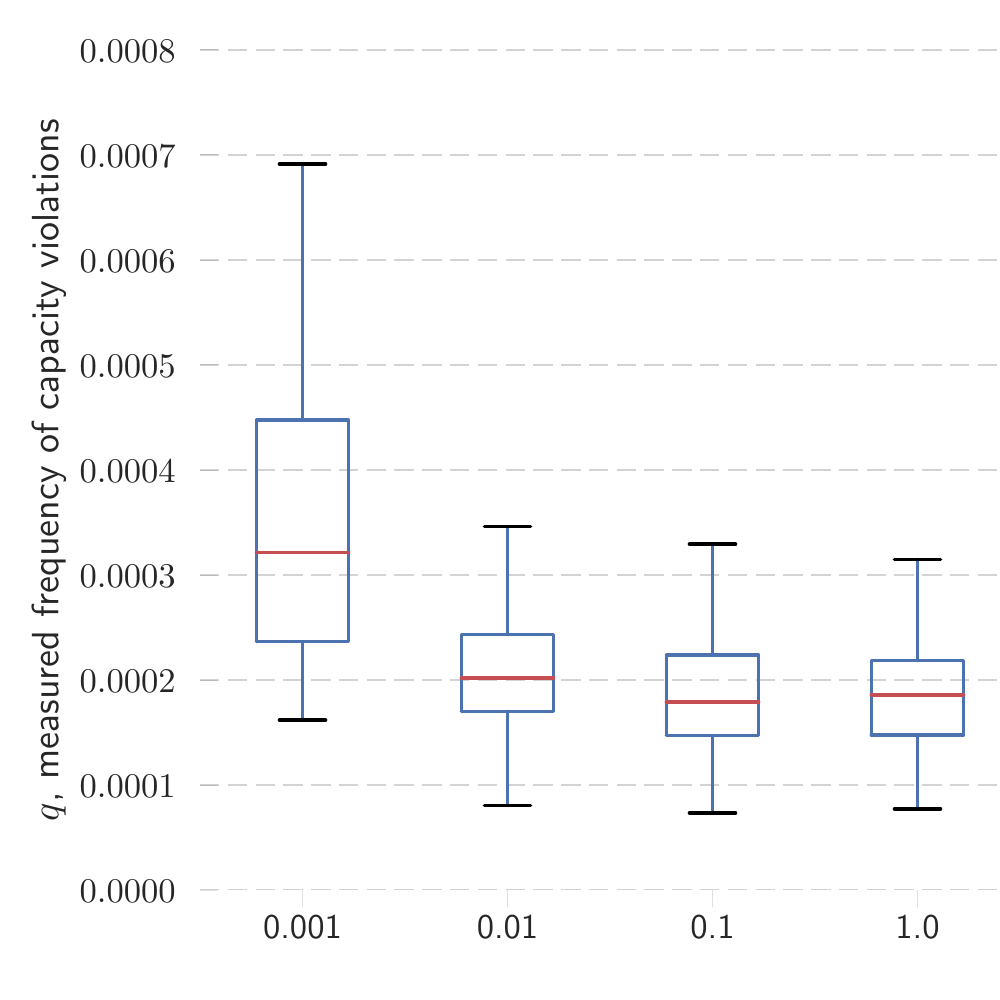}}}%
\caption{$q$ for GPA and \emph{av} as a function of different clairvoyance levels.$1$ is full clairvoyance; $0.001$ corresponds to 10 observations; $0.01$ to 100, etc. $c=1$.}\label{fig:clairvoyance}
\end{figure}

Next, we analyze how the algorithms are affected by reduced quality of input data. We vary the clairvoyance level, i.e., the fraction of samples belonging to the observation set, in $\{ 0.001, 0.01, 0.1, \allowbreak 0.3,\allowbreak 0.5, \allowbreak 0.8, 1.0 \}$. For instance, for clairvoyance level $0.001$, the estimators have $0.001 \cdot 10\,000$, or just 10 \emph{inst} observations to build the tasks' statistical model; to compute the empirical QoS $q$, the produced packing is then evaluated on the remaining $9\,990$ observations (the only difference is clairvoyance 1.0, for which the estimation and the evaluation sets both consisted of all $R=10\,000$ samples). Figure~\ref{fig:clairvoyance} summarizes the results for \emph{av} (which we treat as a baseline) and GPA estimators. We omit results for other \emph{av}; we also omit results for GPA with other thresholds $\varrho$, as they were similar to $\varrho=0.01$. Figures for GPA and \emph{av} have different Y scales: our goal is to compare the relative differences between algorithms, rather than the absolute values (which we do in Section~\ref{sec:exp-comp-algorithms}).

Just 100 observations are sufficient to achieve a similar empirical QoS level $q$ as the fully-clairvoyant variant for both GPA and \emph{av}. although, comparing results for levels $0.01$ and $0.1$, GPA has a slightly higher mean than \emph{av}.
As we demonstrated in Section~\ref{sec:exp-comp-algorithms}, GPA underestimates rare events; hiding data only magnifies this effect.
Furthermore, as the model build by GPA is more complex (it estimates both the average and the standard deviation),
for smaller clairvoyance levels,
we expected GPA to have relatively worse $q$.
However, with the exception of the smallest threshold $\varrho=0.001$, the relative degeneration of GPA and of \emph{av} is similar. 

\subsection{Impact of Rebalancing on Frequency of Capacity Violations}\label{sec:exp-repackaging}

\begin{figure}[tb]
  \centering
  \includegraphics[width=\columnwidth]{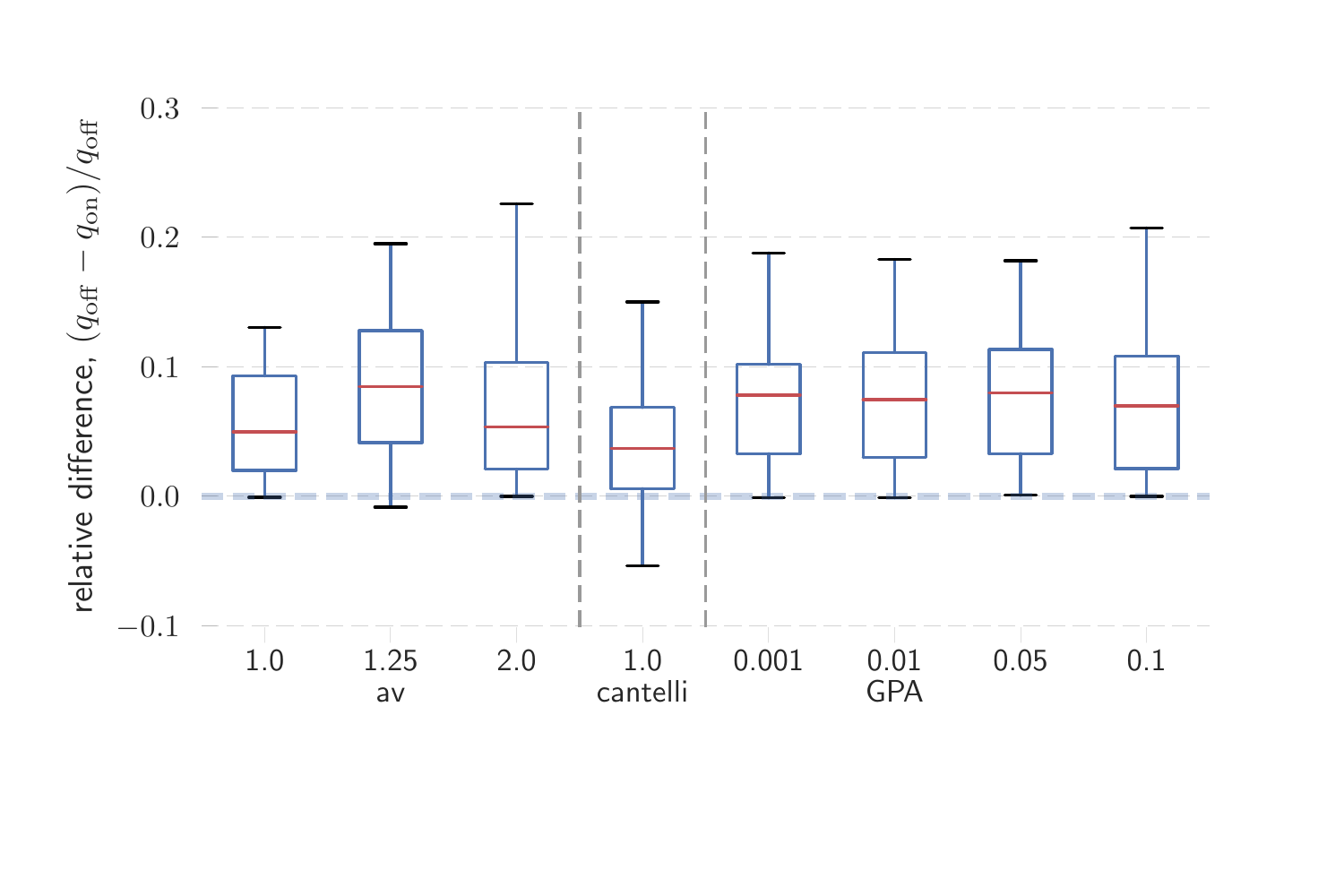}
  \vspace{-5em}\caption{Relative decrease in the measured frequency of capacity violations from rebalancing. $q_{\text{on}}$ denotes $q$ with rebalancing; while $q_{\text{off}}$ without.}\vspace{-1em}
  \label{fig:repacking-gain}
\end{figure}

Finally, we measure how much the rebalancing reduces the frequency of capacity violations,  compared to the results of the on-line bin packing algorithm. Figure~\ref{fig:repacking-gain} shows the \emph{relative} gains achieved by rebalancing (normalized by $q$ of the base algorithm; we omit \emph{perc} and algorithms for which the base algorithm had zero $q$).
Rebalancing uses the unused capacity of the last-opened machine, which is usually severely underloaded, to move some of the tasks from other machines; thus leading to less capacity violations.
The mean relative decrease in capacity violations for both \emph{av} and GPA is around 7\%, which shows that rebalancing modestly improves QoS.
The median absolute number of machines $m_{\text{abs}}$ used by GPA is between 35 (GPA 0.1) and 47 (GPA 0.001).
Thus, the last machine represents at most roughly 2\%-3\% of the overall capacity (in case the machine is almost empty).
As shown in Figure~\ref{fig:exp-gpa-avg-scatter}, the relationship between the capacity and the frequency of capacity violations $q$ is exponential: small capacity increases result in larger decreases of capacity violations.


\section{Related Work}\label{sec:rel-work}
This paper has two principal contributions: the analysis of a new data set of the instantaneous CPU usage; and GPA, a new method of allocating tasks onto machines.

In our data analysis of the Google cluster trace~\cite{clusterdata:Wilkes2011, clusterdata:Reiss2011} (Section~\ref{sec:data-analysis}) we focused on the new information, the instantaneous CPU usage. \cite{clusterdata:Reiss2012b,clusterdata:Di2013} analyze the rest of the trace; and~\cite{carvalho2014longterm} analyzes a related trace (longer and covering more clusters).

Data~center and cloud resource management is an active research area in both systems and theory. A recent survey concentrating on virtual machine placement and on theoretical/simulation approaches is~\cite{pietri_mapping_2016}. Our paper modeled a problem stemming from placement decisions of a data~center scheduler, such as Borg~\cite{verma2015large}; we did not consider many elements, including handling IO~\cite{gog_firmament:_2016,chowdhury_efficient_2015} or optimization towards specific workloads, such as data-parallel/map-reduce computations~\cite{boutin_apollo:_2014,delgado_job-aware_2016}.

We concentrated on bin packing, as our goal was to study how to maintain an SLO when tasks' resource requirements change. If some tasks can be deferred, there is also a scheduling problem;
if the tasks arrive and depart over time, but cannot be deferred, the resulting problem is  dynamic load balancing~\cite{li_dynamic_2014,coffman_dynamic_1983}. We considered simple bin packing as the core sub-problem of these more complex models, as one eventually has to solve packing as a sub-problem. Moreover, scheduling decisions in particular are based on complex policies which are not reflected in the trace and thus hard to model accurately. 

Our method, GPA, uses a standard bin packing algorithm, but changes the fitting criterion.
Bin packing and its variants have been extensively used as a model of data center resource allocation.
For instance,~\cite{DBLP:conf/ipps/TangLRC16} uses a dynamic bin packing model (items have arrival and departure times) with items having known sizes.
\cite{song2014adaptive} studies relaxed, on-line bin packing: they permit migrations of items between bins when bins become overloaded.
Our focus was to model uncertainty of tasks' resource requirements through 
stochastic bin packing.
Theoretical approaches to stochastic bin packing usually solve the problem for jobs having certain distribution. 
\cite{wang_consolidating_2011,breitgand_improving_2012}~consider bin packing with Gaussian items; \cite{zhang_sla_2014}~additionally takes into account bandwidth allocation. \cite{goel1999stochastic}~considers load balancing, knapsack and bin packing with Poisson, exponential and Bernoulli items.
\cite{kleinberg_allocating_2000} for bin packing shows an approximation algorithm for Bernoulli items.
\cite{chen2011effective}~solves the general problem by deterministic bin packing of items; item's size is derived from the item's stochastic distribution (essentially, the machine capacity is divided by the number of items having this distribution that fit according to a given SLO) and correlation with other items. According to their experimental evaluation (on a different, not publicly-available trace, and using 15-minute usage averages), this method overestimates the QoS (for target $\rho=0.05$, they achieve $q=0.02$); they report the number of machines 10\% smaller than \emph{perc 95} (although the later result is in on-line setting: usage is estimated from the previous period).

GPA estimates machine's CPU usage by a Gaussian distribution following the central limit theorem (CLT), perhaps the simplest possible probabilistic model. The CLT has been applied to related problems, including estimation of the completion time of jobs composed of many tasks executed across several virtual machines~\cite{yeo_using_2011}. The stochastic bin packing algorithms that assume Gaussian items proposed for bandwidth consolidation~\cite{wang_consolidating_2011,breitgand_improving_2012} can be also interpreted as a variant of CLT if we drop the Gaussian assumption.



\cite{hwang_hierarchical_2016} addresses stochastic bin packing assuming items' means and standard deviations are known. It essentially proposes to rescale each item's size according to Cantelli inequality (item's mean plus 4.4 or 1.7 times the standard deviation, see Section~\ref{sec:sim-method}). Our experimental analysis in Section~\ref{sec:exp-comp-algorithms} shows that such rescaling overestimates the necessary resources, resulting in allocations using 2.5-5 times more machines than the lower bound. Consequently, for the target 95\% SLO, Cantelli produces QoS of 99.9998\%.

To model resource heterogeneity, bin packing is extended to vector packing: an item's size is a vector with dimensions corresponding to requirements on individual resources (CPU, memory, disk or network bandwidth)~\cite{stillwell2012virtual,lee_validating_2011}.
Our method can be naturally extended to multiple resource type: for each type, we construct a separate GPA; and a task fits into a machine only if it fits in all resource types.
This baseline scenario should be extended to balancing the usage of different kinds of resources~\cite{lee_validating_2011}, so that tasks with complementary requirements are allocated to a single machine.

Our method estimates tasks' mean and standard deviation from tasks' observed instantaneous usage.
\cite{clusterdata:Iglesias2014:task-estimation} combines scheduling and two dimensional bin packing to optimize tasks' cumulative waiting time and machines' utilization. The method uses machine learning to predict tasks' peak CPU and memory consumption based on observing first 24 hours of task's resource usage. If machine's capacity is exceeded, a task is evicted and rescheduled.
While our results are not directly comparable to theirs (as we do not consider scheduling, and thus evictions), we are able to get sufficiently accurate estimates using simpler methods and by observing just 100 samples (Section~\ref{sec:exp-clairvoyance}). While to gather these 100 samples we need roughly 42 trace hours, a monitoring system should be able to take a sufficient number of samples in just a few initial minutes.
\cite{clusterdata:Caglar2014} uses an artificial neural network (ANN) in a combined scheduling and bin packing problem. The network is trained on 695 hours of the Google trace to predict machines' performance in the subsequent hour. Compared to ANN, our model is much simpler, and therefore easier to interpret; we also need less data for training.
\cite{di2015optimization} analyzes resource sharing for streams of tasks to be processed by virtual machines. Sequential and parallel task streams are considered in two scenarios. When there are sufficient resources to run all tasks, optimality conditions are formulated. When the resources are insufficient, fair scheduling policies are proposed. 
\cite{bobroff_dynamic_2007} uses statistics of the past CPU demand of tasks (CDF, autocorrelation, periodograms) to predict the demand in the next period; then they use bin packing to  minimize the number of used bins subject to a constraint on the probability of overloading servers. Our result is that a normal distribution is sufficient for an accurate prediction of a high percentile of the \emph{total} CPU usage of a group of tasks (in contrast to individual task's).


\section{Conclusions}
We analyze a new version of the Google cluster trace that samples tasks' instantaneous CPU requirements in addition to 5-minute averages reported in the previous versions.
We demonstrate that changes in tasks' CPU requirements are significantly higher than
the changes reported by 5-minute averages.
Moreover, the distributions of CPU requirements vary significantly across tasks.
Yet, if ten or more tasks are colocated on a machine, high percentiles of their total CPU requirements can be approximated reasonably well by a Gaussian distribution derived from the tasks' means and standard deviations. However, 99th and 99.9th percentiles tend to be underestimated by this method.
We use this observation to construct the Gaussian Percentile Approximation estimator for stochastic bin packing.
In simulations, GPA constructed colocations with the observed frequency of machines' capacity violations similar to the requested SLO. Nevertheless, because of using the Gaussian model, GPA underestimates rare events: e.g., for a SLO of 0.0010, GPA achieves frequency of 0.0016. Thus, for such SLOs, GPA should be invoked with lower goal tresholds.
Compared to a recently-proposed method based on Cantelli inequality~\cite{hwang_hierarchical_2016}, for 95\% SLO, GPA reduces the number of machines between 1.9 (when Cantelli assumes Gaussian items) and 3.7 (for general items) times.
GPA also turned out to work well with machines with different capacities. Moreover, as input data it requires only the mean and the standard deviation of each task's CPU requirement --- in contrast to the complete distribution. We also demonstrated that these parameters can be adequately estimated from just 100 observations.
Apart from the rebalancing step, our algorithms are on-line: once a task is placed on a machine, it is not moved. Thus, the algorithms can be applied to add a single new task to an existing load (if all tasks are released at the same time). 

Using the Gaussian distribution is a remarkably simple approach---we achieve satisfying QoS  without relying on machine learning~\cite{clusterdata:Iglesias2014:task-estimation} or artificial neural networks~\cite{clusterdata:Caglar2014}. We claim that this proves how important   high-frequency data is for allocation algorithms.

Our analysis can be expanded in a few directions.
We deliberately focused on a minimal algorithmic problem with a significant research interest in the theoretical field --- stochastic bin packing --- but using realistic data. We plan to extend our experiments to stochastic processes (raw data from the trace, rather than stationary distributions generated from them) to validate whether the algorithms still work as expected. We also plan to drop assumptions we used in this early work: to extend packing algorithms to multiple dimensions; to measure and then cope with correlations between tasks; or to pack pools of machines with different capacities.
An orthogonal research direction is a requirement estimator more robust than GPA, as GPA systematically underestimates rare events (small machines or SLOs of 99th or 99.9th percentile).

\begin{acks}
  We thank Jarek Kuśmierek and Krzysztof Grygiel from Google for helpful discussions; Krzysztof Pszeniczny for his help on statistical tests; and anonymous reviewers and our shepherd, John Wilkes, for their helpful feedback. We used computers provided by ICM, the Interdisciplinary Center for Mathematical and Computational Modeling, University of Warsaw.
  The work is supported by 
\grantsponsor{}{Google}{http://dx.doi.org/10.13039/100006785} under Grant
  No.:~\grantnum{GS100006785}{2014-R2-722} (PI: Krzysztof Rzadca).
\end{acks}

\bibliographystyle{ACM-Reference-Format}
\bibliography{ref}

\end{document}